%% file: paper.tex
\documentclass[reprint,aps,pra,twoside,superscriptaddress]{revtex4-1}

\usepackage{hyperref}
\hypersetup{colorlinks=true, linkcolor = blue, citecolor = blue, urlcolor = blue}
\makeatletter\@input{hidelinks}\makeatother 

\usepackage{amsmath,amssymb}
\usepackage{bbold}
\usepackage{units}

\usepackage{dcolumn}

\usepackage{graphicx}

\input{poznachennya.tex}
\input{fizychni_komandy.tex}

\input{dlya_tablychok.tex}

\newcommand{\at}[2]{{#1}\big|_{#2}}
\newcommand{\intR}{\int_{-\infty}^{+\infty}}
\newcommand{\Fourier}{\mathcal F}
\newcommand{\unity}{\mathbb 1}

\DeclareMathOperator{\Real}{Re}

\begin{document}

\title{Single-photon limit of dispersive readout of a qubit with a photodetector}
\author{Andrii~M.~Sokolov}
\email[E-mail: ]{andriy145@gmail.com}
\affiliation{Institute of Physics of the National Academy of Sciences, pr. Nauky 46, Kyiv 03028, Ukraine}
\affiliation{Theoretical Physics, Saarland University, 66123 Saarbr{\"u}cken, Germany}
\author{Eugene~V.~Stolyarov}
\affiliation{Institute of Physics of the National Academy of Sciences, pr. Nauky 46, Kyiv 03028, Ukraine}

\begin{abstract}
We study the dispersive readout of a qubit in the ultimate limit of a single-photon probe. The use of a single-
photon probe avoids the errors due to nonorthogonality of coherent states. A photodetector is used in the scheme
we consider. The dynamics of the system is studied using the Heisenberg-Langevin equations. We treat the
counter-rotating terms in the Hamiltonian perturbatively, which leads to the Bloch-Siegert shift in the resonator
frequency. It is shown how this can improve the readout. The theory of photon transport through the qubit and the
resonator it couples to is provided while taking the effect of the counter-rotating terms into account. To calculate
the readout contrast, we use two approaches. The first one neglects the qubit relaxation and allows us to derive a
compact expression for the contrast. Also, we obtain simple estimates for the system parameters to maximize the
contrast. The second approach accounts for the qubit relaxation, which allows us to further improve the contrast.
We demonstrate that for a readout time of 1$\mu$s, a contrast of more than 75\% can be achieved for an ideal detector and single-photon source.

\end{abstract}

\maketitle

\section{Introduction}

Dispersive measurement~\cite{blais2004cavity,walter2017rapid} is an established method for readout of a superconducting qubit~\cite{clarke2008superconducting,berman2009measurement}.
In the dispersive readout, a qubit is weakly coupled to a resonator.
Depending on the qubit state, the cavity resonance is shifted either to the blue or to the red side.
To probe the cavity, homodyne detection is usually used.
When the cavity is probed with a resonant coherent signal, it acquires a phase shift that depends on the qubit state.
This shift is measured by a homodyne after several amplification stages.
To approach quantum-limited amplification, parametric amplifiers~\cite{roy2018quantum} are used.
This requires additional circulators and drive tones in the cryostat, which makes the setup hard to scale with the number of qubits.

An alternative approach is to use a photodetector~\cite{govia2014high,sokolov2016optimal}.
Suppose the probe frequency is chosen at the cavity resonance for a particular qubit state.
Depending on the state of the qubit, the radiation either mostly passes through the cavity or reflects off it.
A photodetector on the cavity output port provides a click for a particular qubit eigenstate.
The click can be picked up by room-temperature electronics, with no need for a complex and bulky amplification chain~\cite{govia2014high}.
The photodetector scheme was demonstrated in Ref.~\cite{opremcak2018measurement}.

A coherent probe is used in most of the implemented and proposed readout schemes.
States of the output radiation---different for the qubit in the ground and the excited states---are approximately coherent, too.
As coherent states are non-orthogonal, it is impossible to discriminate them without errors.
This contributes to the readout error.
To overcome this, in Refs.~\cite{divincenzo2014dispersive,didier2015heisenberg} the homodyne readout is modified as the output radiation is squeezed.
However, even more circulators and drives are needed in the input chain.
The proposed protocols make the homodyne measurement even harder to scale.

A Fock-state probe in the photodetector scheme can be used to avoid the errors due to the non-orthogonality.
In this work, we study the dispersive readout with the smallest possible amount of energy---a single photon---probing the resonator.
That is also the simplest case for both experiment and theory.
A single-photon pulse can be generated with a single element: an artificial atom decaying into a waveguide~\cite{peng2016tuneable}.
A simple on--off detector such as a Josephson photomultiplier~\cite{chen2011microwave,opremcak2018measurement}, which can only distinguish a vacuum input state, is suitable for the measurement.
A theoretical description that takes into account the excitation exchange between the qubit and the resonator takes a considerably simpler form for the case of a single input photon.
Some of the results for that case might be helpful for approximate analysis of a more sophisticated case of a multi-photon input pulse.
We expect that the scheme studied can be scaled reasonably well.
Indeed, compared to the readout method reported in Ref.~\cite{opremcak2018measurement}, our scheme only requires an additional circulator.

It is challenging to perform a readout with a single photon.
We enhance the readout efficiency by increasing the qubit-resonator coupling.
With other parameters unchanged, this increases an unwanted exchange of excitations between the qubit and the resonator.
To suppress it, the qubit-resonator detuning $\omega_\q - \omega_\res$ should also be increased.
Eventually, $\omega_\q - \omega_\res$ becomes of the same order of magnitude as $\omega_\q + \omega_\res$, which invalidates the rotating-wave approximation~(RWA).
To remedy this, the counter-rotating terms in the Hamiltonian can be treated in the first order of perturbation theory.
This gives a Bloch-Siegert shift in the cavity resonance~\cite{zueco2009qubit,beadoin2011dissipation} as demonstrated in the experiment of Ref.~\cite{forndiaz2010observation}.
We show how this shift can be used to improve readout.

\begin{figure*}[t!]
	\includegraphics{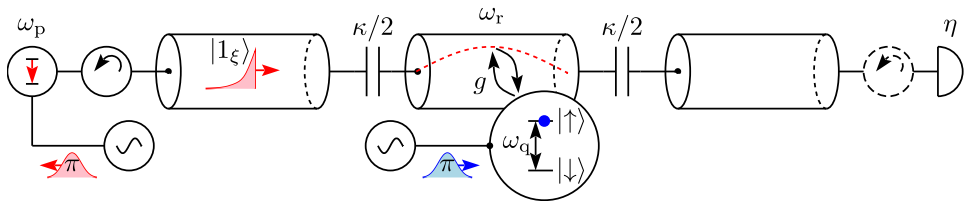}
	\caption{Measurement scheme.
		A single-photon pulse $\ket{1_\xi}$ of exponentially damped shape is incident on the first port of a cavity.
		The cavity resonant frequency is $\omega_\res$ and its leakage is $\kappa/2$ to each of its ports.
		The cavity is dispersively coupled to a qubit.
		The qubit transition frequency is $\omega_\q$.
		It is prepared at the initial moment of time.
		To prepare the excited state $\ket \uparrow$, a $\pi$ pulse is used.
		On the second port there is an on--off photodetector with a quantum efficiency $\textstyle \eta$.
		The pulse can be generated by decay of a two-level system with the transition frequency $\omega_\ph$.
		The two-level system is protected from a reflected photon by a circulator.
		The dashed circulator indicates there is no back-action on the cavity due to reflection off the detector.
		The back-action can also be avoided without a circulator~\cite{opremcak2018measurement}.
	}
	\label{figScheme}
\end{figure*}

To assess the readout performance, one needs a theory of transport of Fock-state photons through the resonator and the qubit beyond the RWA.
The previous studies of the few-photon transport through the resonator coupled to a qubit (see, e.g., Refs. \cite{shen2009theory,rephaeli2012few,oehri2015tunable,hu2018,stolyarov2019few}) use the RWA to describe the resonator-qubit interaction. 
In this work, we go beyond the RWA in treating the photon transport by systematically taking into account the Bloch-Siegert shift.

Dissipation to the waveguides of the qubit and the resonator beyond the RWA should be treated carefully~\cite{beadoin2011dissipation}.
While a suitable master equation for that case is derived in that reference, we choose a different method to treat the system.
Our treatment of the system dynamics and photon transport is based on the Heisenberg-Langevin equations for the entire system including the waveguides.
The approach is inspired by Refs.~\cite{chumak2013phase,berman2011influence}.
It allows us to highlight some subtle moments that arise due to the break of the RWA.
While we focus on the superconducting qubit readout, our treatment is applicable to other types of qubits that couple strongly to a cavity.

The paper is organized as follows.
We review the readout scheme we study in Sec.~\ref{secMeasurementScheme}.
In Sec.~\ref{secHamiltonian} we write the Hamiltonian of the system.
A theory of photon transport that neglects the qubit relaxation is developed in Sec.~\ref{secTransport}.
Using this approach, we derive a simple expression for the contrast in Sec.~\ref{secContrast}, and we estimate the readout performance in Sec.~\ref{secEstimates}.
Section~\ref{secRelaxation} presents the results obtained using the approach that accounts for the qubit relaxation.
We discuss and sum up our results in Sec.~\ref{secDiscussionOutlook}.
The additional discussions and detailed derivations are delegated to Appendices.

\section{Measurement scheme}
\label{secMeasurementScheme}

The readout setup is schematically depicted in Fig.~\ref{figScheme}.
The resonance $\omega_\res$ of the cavity is shifted to $\omega_\res + \chi$ for the excited qubit state $\ket \uparrow$, and $\omega_\res - \chi$ for its ground state $\ket \downarrow$.
At one of the resonances, a probe photon is incident on the cavity.
Suppose the photon central frequency is $\omega_\ph = \omega_\res + \chi$.
If the qubit is in the excited state, it is most likely that the photon passes through the cavity.
Then the detector delivers a click, which indicates that the qubit is excited.
If there is no click, we decide that the qubit is in the ground state.
Due to a large qubit-resonator detuning, it is unlikely that they exchange an excitation.
Hence the measurement scheme can be highly quantum-non-demolition~\cite{braginsky1996quantum}.

In what follows, we use the following convention on the measurement sequence.
At $t=0$ the probe photon is far from the resonator, so its influence on the cavity, the qubit, and the detector is negligible.
The pulse front reaches the cavity port at $t=t_0$.
One waits for the detector clicks from $t=0$ to $\tm$, where $\tm$ is referred to as the measurement time.

To characterize the readout scheme, one needs a theory of single-photon transport through the resonator-qubit system.
In the next two sections, we develop such a theory.

\section{Hamiltonian}
\label{secHamiltonian}

\begin{figure}[b!]
\includegraphics{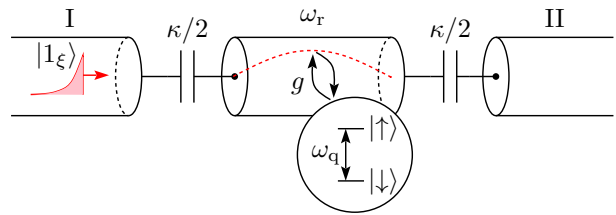}
\caption{The system as modeled by the Hamiltonian.
The semi-infinite waveguides $\bi$ and $\bii$ couple to the resonator.
They model the lack of back action on the cavity and the qubit:
after a photon scatters off the resonator, it does not return back.
Notations are as in Fig.~\ref{figScheme}.}
\label{figModel}
\end{figure}

We model the system as shown in Fig.~\ref{figModel}.
The system Hamiltonian reads
\begin{equation}
\label{eqHamiltonian}
	H = H_\q + H_{\q\res} + H_\res
		+ H_{\res\bi} + H_{\res\bii} + H_\bi + H_\bii,
\end{equation}
where
\begin{gather}
\label{eqHqHr}
	H_\q = \frac 1 2 \hbar \omega_\q \sigmaz,
\quad
	H_\res = \hbar \omega_\res \left(a^\dag a + \frac 1 2\right),
\\
\label{eqHqr}
	H_{\q\res} = \hbar g \sigmax (a + a^\dag)
\end{gather}
are the Hamiltonians of the qubit, the resonator, and the qubit-resonator interaction, respectively.
Together, these three comprise the Rabi Hamiltonian.
Here and below we use the following notations:
$\sigmax$, $\sigmay$, and $\sigmaz$ for the Pauli operators of the qubit (quasi)spin;
$\sigma_\pm = \frac 1 2 (\sigmax \pm i \sigmay)$ for the raising and lowering operators of the qubit;
$a^\dag$ and $a$ for the resonator photon creation and annihilation operators;
$g$ for the qubit-resonator coupling strength.
Hamiltonians of the waveguide fields are
\begin{equation}
\label{eqHw}
	H_\alpha = \hbar \int_0^\infty dk \omega_k b_k^{\alpha\dag} b^\alpha_k,
\quad
	\alpha = \bi, \bii,
\end{equation}
where the contribution of the zero-point oscillations is omitted.
$b_k^{\alpha\dag}$ and $b^\alpha_k$ are the operators of creation and annihilation of a photon with wave vector $\textstyle k$ and frequency $\textstyle \omega_k$ in the waveguide with index $\textstyle \alpha$.
For both waveguides the linear dispersion relation
\begin{equation}
\label{eqDispersion}
	\omega_k = v k
\end{equation}
holds, where $v$ is the velocity of propagating photons.
The term responsible for the interaction of the waveguides with the resonator is
\begin{equation}
\label{eqHwr}
	H_{\res\alpha} = i \hbar \int_0^\infty dk f_k
		(a - a^\dag)(b^\alpha_k + b_k^{\alpha\dag}).
\end{equation}
We assumed that the resonator couples equally to both waveguides.

The choice of signs in the Hamiltonians~\eqref{eqHqr} and~\eqref{eqHwr} capture the case of different-type couplings: capacitive waveguide-resonator coupling and inductive qubit-resonator interaction, or vice versa.
If we don't drop the fast-oscillating terms in the Hamiltonians, they are not equivalent to those with same-type couplings~\cite{bambaogawa2014recipe}.

However, the change of the coupling type does not alter the main results of this paper.
Appendix~\ref{apOtherCouplings} outlines changes in the case of a general linear transversal coupling.
A partial case of a same-type interaction is discussed there as well.

\subsection{Bloch-Siegert regime}

The case of the qubit strongly detuned from the resonator is of interest.
If $\omega_\q -\omega_\res \sim \omega_\q + \omega_\res$, the RWA breaks down.
We assume, however, that the frequencies are of the same order of magnitude,
\begin{equation}
\label{eqWqWrSameOrderOfMagnitude}
\omega_\q \sim \omega_\res,
\end{equation}
and there is a small parameter
\begin{equation}
\label{eqBlochSiegert}
	\Lambda^2 \ll 1, \quad \Lambda = g / (\omega_\q + \omega_\res).
\end{equation}
We also assume that there are not more than two photons in the cavity: one photon can come from the single-photon input pulse and another one can appear due to the excitation exchange with the qubit.
Under these assumptions, the terms $\propto (a^\dag\sigma_+ + a\sigma_-)$ in the Rabi Hamiltonian can be treated as a perturbation and eliminated via the unitary transformation
\begin{equation}
 \label{eqBlochSiegertTransform}
  \UBS = \exp(\Lambda a\sigma_- - \Lambda a^\dag \sigma_+).
\end{equation}
Transforming the Rabi Hamiltonian~(\ref{eqHqHr}) and (\ref{eqHqr}) with $\textstyle \UBS$ gives
\begin{align}
\label{eqHJC}
	H_\q &+ H_{\q\res} + H_\res
		\to \UBS^\dag(H_\q + H_{\q\res} + H_\res)\UBS
\nonumber \\
		&= H_\q + \BS H_{\q\res} + \BS H_\res + O(\Lambda^2),
\\
\label{eqHqrNoCounterRotating}
	\BS H_{\q\res} &= \hbar g (\sigma_+a + a^\dag \sigma_-),
\\
\label{eqHrBlochSiegert}
	\BS H_\res &= \hbar(\omega_\res + g\Lambda\sigmaz)
						\left(a^\dag a + \frac 1 2\right).
\end{align}
The shift $g\Lambda\sigmaz$ in the cavity resonance is known as the Bloch-Siegert shift~\cite{beadoin2011dissipation}.

We have omitted $\hbar g\Lambda \sigmaz a^2$ and its conjugate in Eq.~\eqref{eqHJC}.
Upon integration of the equations of motion for $\sigma_\pm$ and $a^{(\dag)}$, these terms contribute in the order of $g\Lambda/\omega_\res$.
Due to the condition~\eqref{eqWqWrSameOrderOfMagnitude}, this is of order $\Lambda^2$ and should be neglected.
In the same approximation, the transform~\eqref{eqBlochSiegertTransform} is identical to that used in Ref.~\cite{beadoin2011dissipation}.

Now we transform the rest of the terms in the full Hamiltonian~\eqref{eqHamiltonian}.
Using Eqs.~(\ref{eqHJC})--(\ref{eqHrBlochSiegert}) and $a \to \UBS^\dag a \UBS	= a - \Lambda\sigma_+ + O(\Lambda^2)$,
one gets
\begin{equation}
\label{eqBSHamiltonian}
\begin{split}
	H \to \BS H = &H_\q + H_{\q\res} + \BS H_\res
		+ H_{\res\bi} + H_{\res\bii}
\\
		&+ H'_{\q\bi} + H'_{\q\bii}
		+ H_\bi + H_\bii
		+ O(\Lambda^2).
\end{split}
\end{equation}
The term
\begin{equation}
\label{eqHwqBlochSiegert}
	\BS H_{\q\alpha} = \hbar \Lambda \int_0^\infty dk f_k
		\sigmay (b^\alpha_k + b^{\alpha\dag}_k)
\end{equation}
describes the direct coupling between the dressed qubit and the waveguide.
The Hamiltonian~\eqref{eqBSHamiltonian} allows for Purcell decay in which the qubit relaxes to the waveguides via the resonator.
To take account of the Purcell decay, the Hamiltonian~\eqref{eqBSHamiltonian} is used in Appendix~\ref{apBeyondDispersive} to model the single-photon transport through the resonator-qubit system.

In the next two sections, we assume that the qubit relaxation is negligible.
It is possible to obtain analytical results for that case.

\subsection{Dispersive Bloch-Siegert picture}

The Hamiltonian~(\ref{eqHJC})--(\ref{eqHrBlochSiegert}) of the qubit-resonator subsystem is of the Jaynes-Cummings form.
It is possible to diagonalize it with a treatment similar to that of Ref.~\cite{blais2004cavity}.
The resonator-qubit detuning is large,
\begin{equation}
\label{eqDispersiveRegime}
	4 \lambda^2 \ll 1, \quad \lambda = g / (\omega_\q - \omega_\res).
\end{equation}
That is the critical photon number criterion from Refs.~\cite{blais2004cavity,boissoneault2009dispersive} for a cavity populated by a single photon.
The photon comes from the input pulse.
The dispersive approximation is valid under the condition~\eqref{eqDispersiveRegime}, and we neglect the possibility that the qubit will provide another photon.
As $|\lambda| > \Lambda$, the inequality~\eqref{eqBlochSiegert} follows from the last one, and $O(\Lambda^2)+ O(\lambda^2) = O(\lambda^2)$ as well as $O(\Lambda\lambda) = O(\lambda^2)$.
The dispersive transform
\begin{equation}
\label{eqDispersiveTransform}
	\Udisp = \exp(-\lambda \sigma_+ a + \lambda a^\dag \sigma_-)
\end{equation}
then approximately diagonalizes the Hamiltonian~(\ref{eqHJC}).
Applying the transform yields
\begin{align}
\label{eqDressedBSHamiltonian}
\begin{split}
	H &\to H_\q + \BSd H_\res
\\
		&+ H_{\res\bi} + H_{\res\bii}
		+ \BSd H_{\q\bi} + \BSd H_{\q\bii}
\\
		&+ H_\bi + H_\bii + O(\lambda^2),
\end{split}
\\
\label{eqHrDressed}
	\BSd H_\res
		&= \hbar(\omega_\res + g\Lambda\sigmaz + g\lambda\sigmaz)
					\left(a^\dag a + \frac 1 2\right),
\\
\label{eqHwq}
	\BSd H_{\q\alpha} &= \hbar (\lambda + \Lambda) \int_0^\infty dk f_k
		\sigmay (b^\alpha_k + b^{\alpha\dag}_k).
\end{align}
It was used that
$a \to \Udisp^\dag a \Udisp = a + \lambda \sigma_- + O(\lambda^2)$
and $\sigma_- \to \sigma_- + \lambda a\sigmaz + O(\lambda^2)$.
The total shift in the resonator frequency is identical to that given in Ref.~\cite{zueco2009qubit}.
It sets the performance of a dispersive readout.
As shown in Appendix~\ref{apOtherCouplings}, it does not change when the qubit-resonator and the resonator-waveguide couplings are of the same type.
The Bloch-Siegert shift $g\Lambda$ becomes comparable with the dispersive one $g\lambda$ when $\omega_\q-\omega_\res \sim \omega_\q + \omega_\res$.
Equations~(\ref{eqDressedBSHamiltonian})--(\ref{eqHwq}) constitute the Hamiltonian in the dispersive Bloch-Siegert picture.
That is the picture we use for the analytical treatment in Secs.~\ref{secTransport}--\ref{secEstimates}.

\section{Photon transport}
\label{secTransport}

Here we calculate the density of transmitted photons for a single-photon Fock pulse with a given shape.
First we link the density to the cavity population;
then we express the population in terms of the incoming pulse spectrum.
The result is generalized for an $N$-photon pulse.

\subsection{Density of transmitted photons}

The density of transmitted photons~\cite{chumak2012operator,chumak2013phase} is
\begin{equation}
\label{eqTransmittedDensity}
	\mean{\rho_\trans(x,t)} = \frac 1 {2\pi} \int_0^\infty \int_0^\infty
			dk dl \mean{b^{\bii\dag}_k(t) b^\bii_l(t)} e^{-i(k-l)x},
\end{equation}
where $x > 0$.

From the Hamiltonian~\eqref{eqDressedBSHamiltonian}, one obtains the equations of motion for the annihilation operators of a waveguide photon:
\begin{equation}
 \begin{split}
  \dot b_k & = \frac 1 {i\hbar} [b_k, H]
  \\
  & = -i\omega_k b_k
  + f_k [a - a^\dag - i(\lambda+\Lambda)\sigmay].
 \end{split}
\end{equation}
Their formal solution is given by
\begin{multline}
\label{eqbkFormalIntegral}
	b_k(t) = b_k(0) e^{-i\omega_k t}
		+ f_k \int_0^t dt'e^{-i\omega_k (t-t')}
\\
		 \times	[a - a^\dag - i(\lambda+\Lambda)\sigmay]_{t'}.
\end{multline}
Waveguide indices are omitted for brevity.
The first term on the right-hand side (rhs) of Eq.~\eqref{eqbkFormalIntegral} represents the free-propagating part of the waveguide field and the second one describes the influence of the qubit and the resonator.

Now we derive two useful identities.
Multiplying Eq.~\eqref{eqbkFormalIntegral} by $e^{ikx}$ and integrating over $k$, one obtains
\begin{multline}
\label{eqbkIntegratedRaw}
	\int_0^\infty dk b_k(t)e^{ikx} 	= \int_0^\infty dk e^{-ikv(t-x/v)} b_k(0)
\\
		+ \int_0^t dt' \int_0^\infty dk f_k e^{-ikv(t-t'-x/v)}
\\
		 \times	[a - a^\dag - (\lambda+\Lambda)(\sigma_+ - \sigma_-)]_{t'},
\end{multline}
where the dispersion relation~\eqref{eqDispersion} was used.
Consider the second term on the rhs.
Approximately, $a$, $a^\dag$, $\sigma_-$, and $\sigma_+$ vary as $e^{-i\omega_\res t'}$, $e^{i\omega_\res t'}$, $e^{-i\omega_\q t'}$, and $e^{i\omega_\q t'}$.
We drop the terms proportional to $\sigma_+(t') e^{ikvt'}$ and $a^\dag(t') e^{ikvt'}$ since they oscillate rapidly and vanish after integration over $t'$.
By a similar argument, we can extend the integration by $k$ to $-\infty$.
The remaining parts of the integrand comprise $\sigma_-(t') e^{ikvt'}$ and $a(t') e^{ikvt'}$ and oscillate fast for $k<0$.
Next, due to integration over $t'$, only narrow regions around respective frequencies of $a$ and $\sigma_-$ contribute significantly.
We assume that the coupling strength $f_k$ is approximately constant in these regions.
Extending the integration to $-\infty$ and using that $\intR dk e^{ikx} = 2\pi\delta(x)$ yields
\begin{multline}
\label{eqbkIntegrated}
	\int_0^\infty dk b_k(t)e^{ikx} 	= \int_0^\infty dk e^{-ikv(t-x/v)} b_k(0)
\\
		+ \frac{2\pi}v \theta(t - \frac x v) \theta(x)
			[f_\res a + f_\q (\lambda+\Lambda) \sigma_-]_{t-x/v},
\end{multline}
where $f_{\res,\q} = f(\omega_{\res,\q} / v)$ and
\begin{equation}
\label{eqHeaviside}
	\theta(t) = \begin{cases}
		0 & \text{ for } t < 0,
	\\
		1/2 & \text{ for } t = 0,
	\\
		1 & \text{ for } t > 0
	\end{cases}
\end{equation}
is the Heaviside step function.
Step functions arise due to integration of $\delta(t' - t + x/v)$ from $t'=0$ to $t$.
At time $t$ a point $x$ in a waveguide is influenced by the qubit and the resonator states at time $t-x/v$ due to finite velocity of propagation.
$\theta(t-x/v)$ ensures that the resonator and the qubit do not influence a point $x$ in a waveguide for $t<x/v$.
The reasoning analogous to that used in obtaining Eq.~\eqref{eqbkIntegrated} leads to a similar identity,
\begin{equation}
\label{eqfkbkIntegrated}
 \begin{split}
	\int_0^\infty dk f_k b_k(t)
      = & \, \int_0^\infty dk f_k e^{-ikvt} b_k(0) \\
        & \, + \frac 1 4 \theta(t)
           [\kappa a + \kappa_\q (\lambda+\Lambda) \sigma_-]_t,
 \end{split}
\end{equation}
where
\begin{gather}
\label{eqKappa}
	\kappa = 4\pi f_\res^2 / v,
\quad
	\kappa_\q = 4\pi f_\q^2 / v.
\end{gather}
It will be seen from what follows that $\kappa$ is the total decay rate of the resonator.
$\kappa_\q$ is the decay rate of the resonator as seen by the Purcell decay~\cite{beadoin2011dissipation}.
Taking $\textstyle f_k \propto \sqrt{k}$~\cite{bambaogawa2014recipe}, one arrives at the relation $\kappa_\q = (\omega_\q/\omega_\res) \kappa$, which we use in Appendix~\ref{apBeyondDispersive}.

In deriving Eqs.~\eqref{eqbkIntegrated} and \eqref{eqfkbkIntegrated} we drop the terms under the integral which are proportional to $\sigma_+(t') e^{ikvt'}$ and $a^\dag(t') e^{ikvt'}$.
This relates to the RWA made to write out Eq.~(A3) of Ref.~\cite{beadoin2011dissipation}.
In contrast to the reference, our approach allows making this approximation naturally.

Substituting Eq.~\eqref{eqbkIntegrated} into Eq.~\eqref{eqTransmittedDensity}, one has
\begin{widetext}
\begin{multline}
\label{eqDensityVerbose}
	\mean{\rho_\trans(x,t)}	=
		 \frac 1 {2\pi} \int_0^\infty\!\!\!\!\int_0^\infty \!\!\! dk dl
				\mean{b^{\bii\dag}_k(0) b^\bii_l(0)} e^{-iv(k-l)(t-x/v)}
		+ \frac{\kappa/2}v \mean{a^\dag a}_{t-x/v}
		+ \frac {2\pi f_\q f_\res}{v^2}	(\lambda+\Lambda)
			\big(\mean{a^\dag \sigma_-}_{t-x/v} + \cc \big)
\\
		+ \Big(\frac 1 v
		\Bigmean{[f_\res a^\dag + f_\q (\lambda+\Lambda) \sigma_+]_{t-x/v}
				\int_0^\infty dk e^{-ikv(t-x/v)} b^\bii_k(0)}
			+ \cc\Big) + O(\lambda^2)
\end{multline}
\end{widetext}
for $t > x/v > 0$.

Now we show that only the second term in Eq.~\eqref{eqDensityVerbose} should be retained.
First, we consider the averages that involve $b^\bii_k$.
Both waveguides, the resonator, and the qubit are entangled in the ground state due to the counter-rotating terms like $\sigma_- b^\alpha_k$ and $a b^\alpha_k$ in $\textstyle \BSd H_{\q\alpha}$~\eqref{eqHwq} and $H_{\res\alpha}$~\eqref{eqHwr}.
However, far from overdamping~\cite{forndiaz2016ultrastrong},
\begin{equation}
\label{eqRWA}
	\kappa, \kappa_\q \ll \omega_\res, \omega_\q,
\end{equation}
and for a narrow-band pulse, the system state is approximately separable.
Then the second waveguide state is close to vacuum at $t=0$.
Indeed, the system is thermalized at a low temperature, $k_\mathrm{B} T \ll \hbar\omega_\res$.
In this case, the number of thermal photons in the waveguides and the resonator is negligibly small.
The input pulse has no effect on the second waveguide at $t=0$.
Hence the resonator-waveguide subsystem is in the ground state.
Therefore, the first term in Eq.~\eqref{eqDensityVerbose} vanishes.
So does the term with $\mean{a^\dag(t) b_k^\bii(0)}$ and its conjugate.
Now we treat the qubit-related averages.
We assume that the qubit and the cavity are not initially correlated.
The correlation arises, over the course of time, in the first order of interaction parameters $\lambda$ and $\Lambda$, $\mean{a^\dag\sigma_\pm} = O(\lambda)$.
Then the terms with $\lambda+\Lambda$ in~\eqref{eqDensityVerbose} are of second order in $\lambda$, which is beyond the accuracy of Eq.~\eqref{eqDensityVerbose}.
Thus finally,
\begin{equation}
\label{eqTransmittedDensityInTermsOfPopulation}
	\mean{\rho_\trans(x, t)} = \frac{\kappa}{2v} \mean{a^\dag a}_{t - x/v},
\quad
	t > \frac x v > 0.
\end{equation}
This expression is interpreted as follows.
In a time $\Delta t$, $\kappa \Delta t/2$ photons leak to the waveguide, where they propagate over a distance $v\Delta t$.
The shape of a propagating pulse follows the cavity population dynamics.
The delay $x/v$ is due to finite velocity of propagation $v$.

\subsection{Cavity population}

Using the Hamiltonian~\eqref{eqDressedBSHamiltonian}, one obtains the equation of motion for the resonator variable
\begin{gather}
\label{eqDiffa}
	\dot a(t) = -i \omegaResEff(t) a(t)
		- \sum_{\alpha = \bi,\bii}
			\int_0^\infty dk f_k (b_k^\alpha + b_k^{\alpha\dag})_t
		+ O(\lambda^2),
\\
\label{eqEffectiveWr}
	\omegaResEff(t)
		= \omega_\res + g\Lambda\sigmaz(t) + g\lambda\sigmaz(t).
\end{gather}
Applying Eq.~\eqref{eqfkbkIntegrated} to Eq.~\eqref{eqDiffa} leads to the Heisenberg-Langevin equation for $t \ge 0$
\begin{multline}
\label{eqDiffaLangevin}
	\dot a(t) = \left[-i \omegaResEff(t) - \frac \kappa 2 \right] a(t)
		- \frac \kappa 2 a^\dag(t)
		- \frac {\kappa_\q} 2 (\lambda+\Lambda)\sigmax(t)
\\
		- \sum_{\alpha = \bi,\bii}
			\int_0^\infty dk f_k (b_k^\alpha(0) e^{-ikvt} + \hc)
		+ O(\lambda^2).
\end{multline}
It follows from the equation that $\kappa$ is the decay rate of the resonator.
Equations~\eqref{eqDiffa}--\eqref{eqDiffaLangevin} are correct to the first order in $\lambda$.
This follows from the accuracy of the Hamiltonian~\eqref{eqDressedBSHamiltonian}.

Now we solve Eq.~\eqref{eqDiffaLangevin}.
Since $\dot{\sigma}_\mathrm{z}(t) = O(\lambda)$, the time dependence of $\omegaResEff(t)$ is of the second order in $\lambda$.
This exceeds the accuracy of Eq.~\eqref{eqDiffaLangevin} and should be neglected.
Integrating Eq.~\eqref{eqDiffaLangevin}, one obtains
\begin{multline}
	a(t) = a(0) e^{-(i\omegaResEff + \kappa/2)t}
		 - \int_0^t dt' e^{-(i\omegaResEff + \kappa/2)(t-t')}
\\
\begin{aligned}
			\times \bigg\{
				&\frac \kappa 2 a^\dag(t)
				+ \frac {\kappa_\q} 2
					(\lambda+\Lambda)\sigmax(t')
\\
				&+ \sum_{\alpha = \bi,\bii} \int_0^\infty dk
					f_k [b_k^\alpha(0) e^{-ikv t'} + \hc]
			\bigg\}.
\end{aligned}
\end{multline}
We assume that the coupling is strong and $\kappa_\q \lesssim g$.
Then the integrands proportional to $\kappa_\q/2$ contribute beyond the accuracy of Eq.~\eqref{eqDiffaLangevin}:
\begin{multline}
	\int_0^t dt' e^{-(i\omegaResEff + \kappa/2)(t-t')}
			(\lambda + \Lambda) \frac {\kappa_\q} 2 \sigma_\pm(t')
\\
		\sim \frac{\kappa_\q}g \frac {\lambda g} {\omega_\q \pm \omega_r}
		\lesssim \lambda^2.
\end{multline}
Moreover, the term with $b_k^{\alpha\dag}(0) e^{i(vk + \omegaResEff)t'}$ oscillates fast and becomes negligible after integration over $t'$.
The same holds for the term with $a^\dag(t')e^{i\omega_\res t'}$, the contribution of which is negligible under the condition~\eqref{eqRWA}.
One can also extend the integration over $t'$ to $-\infty$, as for $t' < 0$ the input pulse does not appreciably influence the cavity.
Then carrying out the integration yields
\begin{gather}
\label{eqa}
\begin{split}
	a(t) \approx \, &a(0) e^{-(i\omegaResEff + \kappa/2)t}
\\
		 &- \frac{if_\res}{\sqrt{\kappa/2}}
				\sum_{\alpha = \bi,\bii} \int_0^\infty dk
					\Kernel(vk) b_k^\alpha(0) e^{-ivkt},
\end{split}
\\
\label{eqK}	
 	\Kernel(\omega) = \frac{\sqrt{\kappa/2}}
						   {i(\omegaResEff - \omega) + \kappa/2}.
\end{gather}
It was taken into account that $\textstyle f(\omega/v) \approx f_\res$ in the vicinity of $\omega = \omega_\res \pm g(\lambda+\Lambda)$.

To calculate the cavity population, one can use an expansion of unity in the whole system Hilbert space,
\begin{equation}
\label{eqUnity}
	\unity = \unity^\q \unity^\res \unity^\bi \unity^\bii.
\end{equation}
Here the unity operators of the system parts are:
\begin{equation}
	\unity^\q = \ket\uparrow \bra\uparrow + \ket\downarrow \bra\downarrow
\end{equation}
for the qubit space,
\begin{equation}
	\unity^\res = \sum_{n^\res = 0}^\infty \ket{n^\res}\bra{n^\res}
\end{equation}
for the cavity space, and
\begin{gather}
	\unity^\alpha = \sum_{n=0}^\infty
		\int_0^\infty dk_1	\ldots \int_0^\infty dk_n
			\ket{\w^\alpha_{k_1 \ldots k_n}} \bra{\w^\alpha_{k_1 \ldots k_n}},
\\
	\label{eqWk1tokn}
	\ket{\w^\alpha_{k_1 \ldots k_n}} = \nu^\alpha(k_1, \ldots k_n)
		\prod_{k = k_1, \ldots k_n} b^{\alpha\dag}_k(0) \ket{0^\alpha}
\end{gather}
for the $\alpha$-th waveguide space.
In Eq.~\eqref{eqWk1tokn}, $\nu^\alpha$ is a normalization constant which satisfies $\braket{\w^\alpha_{k_1 \ldots k_n}}{\w^\alpha_{k_1 \ldots k_n}} = 1$.

First we express $\mean{a^\dag a}$ for an arbitrary state of the input pulse.
By insertion of the unity operator one gets
\begin{equation}
\label{eqa1a}
	\mean{a^\dag a} = \bra\psi a^\dag \unity a \ket\psi,
\end{equation}
where
\begin{equation}
\label{eqInitialState}
	\ket\psi = \ket\q \ket{0^\res} \ket{\w^\bi} \ket{0^\bii}
\end{equation}
is the initial state of the entire system.
It is comprised of wavefunctions of the system parts.
$\ket\q$ is the qubit wavefunction and $\ket{0^\res}$ is that of the resonator;
$\ket{\w^\bi}$ is the wavefunction of the first waveguide and $\ket{0^\bii}$ is that of the second one.
As explained in the course of derivation of Eq.~\eqref{eqTransmittedDensityInTermsOfPopulation}, the initial state~\eqref{eqInitialState} can indeed be considered separable.
The resonator and the second waveguide are in the vacuum state initially.
We substitute the unity expansion~\eqref{eqUnity}--\eqref{eqWk1tokn} into Eq.~\eqref{eqa1a} and use Eq.~\eqref{eqa}.
Then, using the initial state of the system~\eqref{eqInitialState}, one arrives at
\begin{multline}
\label{eqPopulationGeneral}
	\mean{a^\dag a}_t = \frac{f_\res^2}{\kappa/2} \sum_{n=0}^\infty
		\int_0^\infty dk_1 \ldots \int_0^\infty dk_n
\\
			\times \bra\q \Big| \int_0^\infty dk
				\bra{\w^\bi_{k_1 \ldots k_n}}
					\Kernel(vk) b^\bi_k(0) e^{-ivkt}
				\ket{\w^\bi}
			\Big|^2 \ket\q.
\end{multline}
It was used that $a(0) \ket{0^\res} = 0$ and $b^\bii(0) \ket{0^\bii} = 0$.
We also employed the property
\begin{align}
	\nonumber
	\sum_{q' = \uparrow,\downarrow} |\bra{q'} \zeta(\sigmaz) \ket\q|^2
		&= |\braket\uparrow\q \, \zeta(1)|^2
			+ |\braket\downarrow\q \, \zeta(-1)|^2
\\
		&= \bra\q \big|\zeta(\sigmaz)\big|^2 \ket\q,
\end{align}
where $\zeta$ is a function of $\sigmaz$.

Now we provide an expression for the population, given the input pulse is in a single-photon Fock state:
\begin{equation}
\label{eq1Fock}
	\ket{\w^\bi} = \ket{1^\bi_\xi}
		\equiv \intR dk \xi'(k) b_k^{\bi\dag}(0) \ket{0^\bi}.
\end{equation}
Here $\xi'(k)$ is the incident pulse spectrum.
That is, $\xi'(k)$ is the amplitude of the probability density of finding a monochromatic photon with a wave vector $k$.
$\intR |\xi'(k)|^2 = 1$ due to normalization.
We assume the pulse to be narrow-band, i.e., its spectral width is much smaller than its central frequency.
Hence the limits of integration in Eq.~\eqref{eq1Fock} were extended to $-\infty$.
Using Eqs.~\eqref{eq1Fock} and~\eqref{eqPopulationGeneral}, one arrives at
\begin{equation}
\label{eqPopulation1Fock}
	\mean{a^\dag a}_t = \bra\q
			\big|\Fourier[\Kernel(\omega)\xi(\omega)](t)\big|^2
		\ket\q.
\end{equation}
The equation is only applicable for $t \ge 0$ due to the original restriction in the Langevin equation~\eqref{eqDiffaLangevin}.
We have defined
$\Fourier[f(\omega)](t)
	 = (2\pi)^{-1/2} \intR d\omega \exp(-i\omega t) f(\omega)$,
where
\begin{equation}
\label{eqSpectrumInTermsOfFrequency}
	\xi(\omega) = \frac{\xi'(\omega/v)}{\sqrt v}.
\end{equation}
The last expression follows from the dispersion relation~\eqref{eqDispersion} and $n'(k) dk = n(\omega) d\omega$ with $n'(k) = |\xi'(k)|^2$ and $n(\omega) = |\xi(\omega)|^2$.
Density of photons in an increment $d\omega$ is the same as for the corresponding increment $dk$.

Due to the linearity of the system in the dispersive approximation, an $N$-photon Fock pulse populates the resonator $N$ times the one-photon pulse.
The expression of the same form was obtained in Ref.~\cite{divincenzo2014dispersive} for a coherent input pulse.
For this generalization to be valid, the condition~\eqref{eqDispersiveRegime} should be changed to take into account that the cavity can accommodate $N$ photons.
Such a condition is given in Refs.~\cite{blais2004cavity,boissoneault2009dispersive}.

Photon transport does not depend on whether the qubit-resonator and the resonator-waveguide couplings are different or of the same type (see Appendix~\ref{apOtherCouplings}).
Indeed, one can show that Eqs.~\eqref{eqTransmittedDensityInTermsOfPopulation}, \eqref{eqPopulationGeneral}, and all the subsequent ones are the same in both cases.

Having obtained a description of photon transport, we can now assess the performance of our readout scheme.

\section{Readout contrast}
\label{secContrast}

In this section, we define the readout contrast and review the related terminology.
Then, an explicit expression for the readout contrast in our measurement scheme is derived.

Probabilistic measurement contrast~\cite{govia2014high} is defined as
\begin{equation}
\label{eqContrast}
	C = P_{\uparrow|\uparrow} - P_{\uparrow|\downarrow}.
\end{equation}
$P_{m|i}$ is the probability of inferring the qubit to be in state $\ket m$ while it is in state $\ket{i}$.
The contrast~\eqref{eqContrast} is sometimes referred to as fidelity~\cite{gambetta2007protocols,walter2017rapid,johnson2011dispersive,sokolov2016optimal}.

It is however more consistent to reserve the term fidelity for the other, albeit related, quantity.
As we prefer it, measurement fidelity is the probability of a correct measurement result%
~\footnote{In our previous work~\cite{sokolov2016optimal} the quantity we call here the measurement contrast was erroneously claimed to equal the probability of correct measurement result.}.
Let $P_\downarrow$ and $P_\uparrow$ denote the probabilities to measure $\ket\downarrow$ and $\ket\uparrow$, respectively.
Then fidelity is
\begin{equation}
	F = 1 - P_\downarrow P_{\uparrow|\downarrow}
		 - P_\uparrow P_{\downarrow|\uparrow}
\end{equation}
We know nothing about the qubit initial state prior to readout.
Hence it is reasonable to set $P_{\downarrow,\uparrow} = 1/2$ and
\begin{equation}
\label{eqFidelity}
	F = 1 - (P_{\uparrow|\downarrow} + P_{\downarrow|\uparrow})/2.
\end{equation}
That is the formula given, for example, in Ref.~\cite{fan2014nonabsorbing}.
Taking into account that $P_{\uparrow|\uparrow} + P_{\uparrow|\downarrow} = 1$, we express fidelity~\eqref{eqFidelity} in terms of the probabilistic contrast~\eqref{eqContrast}:
\begin{equation}
\label{eqFidelityInTermsOfContrast}
	F = (1 + C)/2.
\end{equation}

Next we calculate the contrast in our setup.
The measurement outcome is based on the state of an on--off photodetector on the second cavity port.
We assign the readout result to be ``$\uparrow$'' when there is a click and ``$\downarrow$'' in the other case.
Hence Eq.~\eqref{eqContrast} turns to
\begin{equation}
\label{eqContrastOurScheme}
	C = P_{\click|\uparrow}[\xi(\omega)] - P_{\click|\downarrow}[\xi(\omega)],
\end{equation}
where $P_{\click|q}$ is the probability of a click, given the qubit is prepared in an eigenstate $q = \uparrow, \downarrow$, and the cavity is irradiated by a pulse with spectrum $\xi(\omega)$.

The way we decide on the readout outcome is easy to justify in the dispersive regime when the condition~\eqref{eqDispersiveRegime} holds.
In this regime, the qubit does not decay to the waveguides.
Also, recall the input pulse is single-photon.
Then at most one photon reaches the second waveguide and the detector.
In this case, the probability of a click is
\begin{equation} \label{eqPclick}
 P_\click = \eta \mean{\Ntr},
\end{equation}
where $\textstyle \Ntr = \int_0^\tm dt v \rho_\trans(t)$ is the total number of photons transmitted through the cavity, $\textstyle \tm$ denotes the counting time, and $\textstyle \eta$ is a quantum efficiency of the photodetector.
The expression for the click probability valid for an arbitrary number of photons hitting the detector is given in Appendix~\ref{apBeyondDispersive} [see Eq.~\eqref{eqPcl}], where we are interested in the case when the qubit can contribute an additional photon via the Purcell decay.
With Eq.~\eqref{eqTransmittedDensityInTermsOfPopulation}, Eq.~\eqref{eqPclick} yields
\begin{equation}
\label{eqPclInTermsOfPopulation}
	P_\click = \eta \frac{\kappa}{2} \int_0^\tm dt \mean{a^\dag a}_t.
\end{equation}
Suppose we have a high-Q resonator and a narrow-band incoming pulse.
Let the pulse be in resonance with the cavity if the qubit is excited:
\begin{gather}
	\omega_\ph = \bra\uparrow \omegaResEff \ket\uparrow
			 = \omega_\res + \chi,
\\
\label{eqTotalPull}
	\chi = g(\lambda+\Lambda),
\end{gather}
where $\omegaResEff$ is defined in Eq.~\eqref{eqEffectiveWr} and $\chi$ is the total shift of the cavity resonance.
Then by Eqs.~\eqref{eqPopulation1Fock} and~\eqref{eqK} the resonator reflects most of a pulse if the qubit is in the ground state.
Most likely, the detector does not click in this case.
On the other hand, if the qubit is excited, the resonator transmits most of the pulse to the detector port.
It is most probable then for the detector to deliver a click.

In the dispersive regime Eq.~\eqref{eqContrastOurScheme} simplifies.
Cavity population~\eqref{eqPopulation1Fock} is symmetrical with respect to a qubit flip and a shift of $\xi(\omega)$:
\begin{gather}
	\uparrow\, \to\, \downarrow,
\quad
	\xi(\omega) \to \xi(\omega + 2\chi),
\quad
	\mean{a^\dag a} \to \mean{a^\dag a};
\\
	\downarrow\, \to\, \uparrow,
\quad
	\xi(\omega) \to \xi(\omega - 2\chi),
\quad
	\mean{a^\dag a} \to \mean{a^\dag a}.
\end{gather}
Due to Eq.~\eqref{eqPclInTermsOfPopulation} the symmetry applies to the click probability too.
Hence
\begin{equation}
\label{eqContrastDispersive}
	C = P_{\click|\uparrow}[\xi(\omega)]
		- P_{\click|\uparrow}[\xi(\omega - 2\chi)].
\end{equation}

Consider an exponentially damped pulse~\cite{schoendorf2018optimizing} incident on the resonator~(see Fig.~\ref{figScheme}).
Let the pulse front arrive at the resonator at time $t_0$.
The amplitude of the probability density to find a photon at point $x$ in the first waveguide is
\begin{equation}
	\varrho'(x) = \frac{1}{\sqrt{vt_\ph}}
					\exp\left(i k_\ph x + \frac{x+vt_0}{2vt_\ph}\right)
					\theta(-x-vt_0),
\end{equation}
where $k_\ph = \omega_\ph/v$ with $\omega_\ph$ being the central frequency of the pulse.
The probability density $|\varrho'(x)|^2$ of the pulse decays over the length $vt_\ph$;
hence $t_\ph$ is regarded as the pulse duration.
The pulse spectrum is, up to a non-relevant phase:
\begin{equation}
\label{eqLorentzian}
 \begin{split}
  \xi'(k) & = \Fourier[\varrho'(x)](k) \\
  & = \frac 1 {\sqrt{2\pi v t_\ph}} \frac {e^{ikvt_0}} {k - k_\ph + i (2 v t_\ph)^{-1}}.
 \end{split}
\end{equation}

It is convenient to introduce dimensionless quantities
\begin{gather}
\label{eqDimensionlessTK}
	\tau = \frac{t - t_0}{t_\ph},
\quad
	K = \kappa t_\ph,
\\
\label{eqDimensionlessDX}
	D = (\omega_\res + \chi \sigmaz - \omega_\ph) t_\ph,
\quad
	X = \chi t_\ph.
\end{gather}
Equations~\eqref{eqPopulation1Fock}--\eqref{eqSpectrumInTermsOfFrequency}, upon insertion of Eq.~\eqref{eqLorentzian} yield
\begin{multline}
\label{eqLorentzianPopulation}
	\mean{a^\dag a} = \theta(\tau) \bra q
		\frac{4K e^{-(K + 1)\tau/2}}{(K - 1)^2 + 4 D^2}
\\
		\times \big[\cosh\frac{(K - 1)\tau}2 - \cos D\tau\big] \ket q.
\end{multline}
If $K=1$ and $D=0$, one obtains
\begin{equation}
\label{eqLorentzianPopulationK1D0}
	\mean{a^\dag a} = \theta(\tau) \tfrac 1 2 e^{-\tau} \tau^2.
\end{equation}
Using this result, one can check that for the exponentially damped pulse
\begin{equation}
\label{eqContrastDispersiveLorentzian}
	C = \at{P_\click}{D=0} - \at{P_\click}{D=2X}.
\end{equation}
The contrast is compromised by unwanted scattering.
The first term in Eq.~\eqref{eqContrastDispersiveLorentzian} is less than unity, as a non-monochromatic photon can reflect off the cavity even in resonance.
The second term describes the loss of contrast due to the false photon count.
It occurs when a photon passes the cavity off-resonance.

\begin{figure}[t!]
	\centering
	\includegraphics[width = 0.35\textwidth]{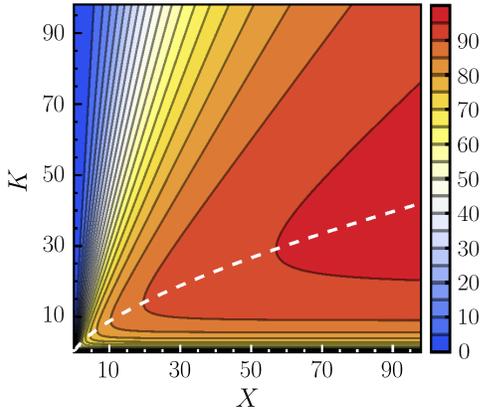}
	\caption{Readout contrast as a function of the dimensionless dispersive pull $X$ and cavity leakage $K$~\eqref{eqDimensionlessDX}.
		The dashed line shows the position of maximum for each $X$.}
	\label{figContrast}
\end{figure}

Suppose the measurement is long enough for the detector to absorb most of the outgoing pulse energy.
That is, $\tm \gg t_\ph, \kappa^{-1}$.
Then one can integrate to $\textstyle \infty$ in Eq.~\eqref{eqPclInTermsOfPopulation}.
Performing the integration using Eqs.~\eqref{eqLorentzianPopulationK1D0} and~\eqref{eqLorentzianPopulation} gives
\begin{align}
	\label{eqnTuned}
	\at{P_\click}{D=0} &= \frac{\eta K}{K+1},
\\
	\label{eqnDetuned}
	\at{P_\click}{D=2X} &= \frac{\eta K(K+1)}{(K+1)^2 + 16X^2}.
\end{align}
Equations~\eqref{eqContrastDispersiveLorentzian}--\eqref{eqnDetuned} and~\eqref{eqDimensionlessTK}--\eqref{eqDimensionlessDX} constitute the expression for the contrast.
The contrast is shown in Fig.~\ref{figContrast}.

Detuning the probe photon from a cavity resonance $\omega_\ph \ne \omega_\res \pm \chi$ lowers the contrast.
To check this, one can straightforwardly generalize the expressions for the contrast for a non-zero detuning.
Note that when a coherent-state probe is used, the maximum of contrast is away from the resonance due to the shot noise~\cite{sokolov2016optimal}.

For a given $X$, the contrast is maximized for
\begin{equation}
	K = u + \frac 1{4u} - \frac 1 2,
\quad
	u = \sqrt[3]{ X\sqrt{16X^2 + 1} + \frac{16X^2 + 1}4 - \frac 1 8}.
\end{equation}
The position of maximum is shown in Fig.~\ref{figContrast}.
For $2X^{2/3} \gg 1$ one obtains that
\begin{equation}
\label{eqOptimalKForBigX}
	K \approx 2X^{2/3}.
\end{equation}
Note it follows that $\chi > \kappa$.

As $C$ grows with $X$, the case of a large dispersive pull is of interest.
In this case, one can give a simple expression for the maximal contrast.
It is shown in Appendix~\ref{apApproximating} that the contrast can be approximated as
\begin{equation}
\label{eqContrastApproximation}
	C \approx \eta\left(1 - \frac {3} {2\kappa t_\ph}\right)
\end{equation}
if the cavity decay rate $K = \kappa t_\ph$ is optimal, as given by Eq.~\eqref{eqOptimalKForBigX}, and
\begin{equation}
\label{eqSimpleContrastCondition}
	X = \chi t_\ph \gtrsim 100,
\end{equation}
with $\chi$ given by Eq.~\eqref{eqTotalPull}.
As follows from the derivation of Eq.~\eqref{eqContrastApproximation}, a third of the contrast loss is due to the false photon count off the resonance.
The other two-thirds are from the absence of a count in the resonance.

Equations~\eqref{eqContrastApproximation} and~\eqref{eqOptimalKForBigX} are the quantitative version of the general considerations given in Ref.~\cite{berman2012dynamics}.
To readout the qubit in our setup means to distinguish a change $2\chi$ in the resonator frequency.
This can only be accomplished if
\begin{equation}
 \label{eqCondDistinguish}
 2\chi\tm > 1,
\end{equation}
where the measurement time $\tm$ is of the same order of magnitude as the pulse duration $t_\ph$.

\section{Estimates}
\label{secEstimates}

Here we use our analytical results to choose the system parameters;
the qubit relaxation is neglected.
The main idea is to relate the minimal measurement time for obtaining a given contrast with the time for the qubit to stay intact.
One also takes care to get acceptable errors due to finite counting time and to avoid qubit relaxation due to the counter-rotating terms in the Hamiltonian.

\subsection{Minimal pulse duration to get a given error}

Here we determine a pulse duration $t_\ph$ that suffices to perform a readout with a given accuracy.
The counting time is considered infinite.

It is convenient to argue in terms of the probability of an erroneous readout
\begin{equation}
\label{eqErrorInTermsOfFidelity}
	\varepsilon = 1 - F.
\end{equation}
By expressing $F$ in terms of contrast with Eq.~\eqref{eqFidelityInTermsOfContrast} and using the approximation~\eqref{eqContrastApproximation} for the latter, one gets
\begin{equation}
	\varepsilon = \frac{1-\eta}2 + \frac{3\eta}{8(\chi t_\ph)^{2/3}}.
\end{equation}
Let us assume $\eta=1$.
Then
\begin{equation}
\label{eqMinTmToGetFixedError}
	t_\ph \ge \frac 1 {\chi} \left(\frac 3 {8\varepsilon} \right)^{3/2}
\end{equation}
suffices to get an error not exceeding $\varepsilon$.

\subsection{Error due to a finite counting time}

Let us calculate the degradation of contrast due to finite counting time.
Integration in Eq.~\eqref{eqPclInTermsOfPopulation} with limits from $t=0$ to $\tm$ gives
\begin{equation}
	\Pcl(\tm) = \Pcl(\infty) - \Delta(\tm),
\end{equation}
where
\begin{widetext}
\begin{equation}
	\Delta(\tm) = \left(
					\frac{Ke^{-\taum} + e^{-K\taum}}{2K}
					+ \frac{2e^{-(K+1)\taum/2}[(K+1)\cos D\taum + D\sin D\taum]}
							{(K+1)^2 + 4D^2}
				\right)
					\frac{2K^2}{(K-1)^2 + 4D^2},
\end{equation}
\end{widetext}
$\Pcl(\infty)$ is the click probability given by Eq.~\eqref{eqnDetuned}, and $\taum = \tm / t_\ph$.
In the spirit of the approximations used to obtain Eq.~\eqref{eqContrastApproximation}, one has
\begin{gather}
\label{eqContrastLossDueToFiniteIntegrationTimeApproximated}
	\at{\Delta(\taum)}{D=0} \approx (1 + 2/K) e^{-\taum},
\\
	\at{\Delta(\taum)}{D=2X} \approx 0.
\end{gather}
Then, from Eqs.~\eqref{eqContrast} and~\eqref{eqDimensionlessDX} it follows that
\begin{equation}
\label{eqContrastLossDueToFiniteIntegrationTime}
	C(\taum) = C(\infty) - \Delta(\taum),
\end{equation}
where $C(\infty)$ is given by Eqs.~\eqref{eqContrastDispersive} and~\eqref{eqnTuned}--\eqref{eqnDetuned}.

To have $\Delta \approx 0.3\%$, one chooses $\tm = 6 t_\ph$.
In comparison, for $\tm = 3 t_\ph$ the degradation in contrast is already around 5\%.
For both cases, one needs $K \gtrsim 10$.

\subsection{Maximal readout duration for the qubit not to relax}

Qubit relaxation time $T_1$ is limited by the time of its Purcell decay $\TPurcell$.
From Refs.~\cite{sete2014purcell,blais2004cavity}, one has
\begin{equation}
\label{eqT1}
	T_1 < \TPurcell,
\quad
	\TPurcell \approx \frac 1 {\kappa \lambda^2}.
\end{equation}
Here we assume that $\kappa_\q \approx \kappa$.

Thus the readout duration is limited by the condition
\begin{equation}
\label{eqRelaxationNegligible}
	\tm \ll \TPurcell
\end{equation}
The ratio $\TPurcell / \tm$ is chosen to avoid significant errors due to the qubit relaxation.
Using Eqs.~\eqref{eqT1}, \eqref{eqDimensionlessTK}, and~\eqref{eqOptimalKForBigX}, one obtains
\begin{equation}
\label{eqMaxTpDueToQubitRelaxation}
	t_\ph < \left(\frac{t_\ph}{2\TPurcell} \right)^{3/2}
			\frac 1 {\lambda^3 \chi}.
\end{equation}
The ratio $\tm / t_\ph$ is chosen to limit the error due to finite integration time.
The error is given by Eqs.~\eqref{eqContrastLossDueToFiniteIntegrationTime} and~\eqref{eqContrastLossDueToFiniteIntegrationTimeApproximated}.

We don't take into account the correction to the qubit relaxation due to the Bloch-Siegert dressing.
This is justified for
\begin{equation}
\label{eqNegligibleNonRWARelaxation}
	\Lambda^2 \ll \lambda^2.
\end{equation}

Combining Eqs.~\eqref{eqMinTmToGetFixedError} and~\eqref{eqMaxTpDueToQubitRelaxation} yields the limit on readout error,
\begin{equation}
\label{eqErrorSetsLambda}
	(1-C)/2 = \varepsilon > \frac{3\TPurcell}{4t_\ph} \lambda^2.
\end{equation}
To express $\varepsilon$ in terms of contrast $C$, Eqs.~\eqref{eqErrorInTermsOfFidelity} and~\eqref{eqFidelityInTermsOfContrast} were used.
A reasonable choice $\tm = \TPurcell/10$ and $\tm = 6t_\ph$ yields $\varepsilon = 45\lambda^2$.

\subsection{Analytics for the parameter choice}
\label{secAnalytics}
Now one can determine all of the system parameters.
By virtue of Eq.~\eqref{eqErrorSetsLambda}, $\lambda$ is set by the readout contrast to be attained.
The other parameters are chosen as follows.
The ratio $\lambda/\Lambda$ is set by the requirement~\eqref{eqNegligibleNonRWARelaxation} which limits the relaxation due to the Bloch-Siegert dressing.
In terms of this ratio,
\begin{equation}
	\omega_\res = \frac{\lambda/\Lambda - 1}{\lambda/\Lambda + 1} \, \omega_\q.
\end{equation}
With this and the definition~\eqref{eqDispersiveRegime} of $\lambda$ one gets
\begin{equation}
\label{eqOptCoupling}
	g = 2\lambda\omega_\q (\lambda/\Lambda + 1)^{-1}.
\end{equation}
Plugging the latter expression into Eq.~\eqref{eqMaxTpDueToQubitRelaxation} and using Eq.~\eqref{eqTotalPull} results in
\begin{equation}
\label{eqFinalTp}
	t_\ph = \frac\lambda{2\Lambda}
			\left(\frac{t_\ph}{2\TPurcell}\right)^{3/2}
			\frac 1 {\lambda^5 \omega_\q}.
\end{equation}
Measurement duration $\tm$ is related to $t_\ph$ by choosing an acceptable error due to finite integration time, which is given by Eqs.~\eqref{eqContrastLossDueToFiniteIntegrationTime} and~\eqref{eqContrastLossDueToFiniteIntegrationTimeApproximated}.
Resonator leakage $\kappa$ can be obtained with Eqs.~\eqref{eqOptimalKForBigX} and~\eqref{eqFinalTp}, and definitions~\eqref{eqDimensionlessTK} and~\eqref{eqDimensionlessDX}.

Equations~\eqref{eqFinalTp} and~\eqref{eqErrorSetsLambda} elucidate what parameters to alter for achieving fast and high-fidelity readout.
Higher $\omega_\q$ is favorable for our readout scheme.
As shown in Ref.~\cite{sokolov2016optimal}, a related scheme with a photodetector also favors higher frequencies.
Higher $\lambda$ is especially beneficial if the qubit does not decay.

The dispersive shift $g\lambda$ and the Bloch-Siegert shift $g\Lambda$ should be of the same sign to maximize the total pull $\chi$~\eqref{eqTotalPull}.
This is the case for $\omega_\q > \omega_\res$.
For $\lambda/\Lambda = 10$ [which satisfies Eq.~\eqref{eqNegligibleNonRWARelaxation}], the pull $\chi$ is about 20\% larger for $\omega_\q > \omega_\res$ than in the opposite case.
Hence $t_\ph$~\eqref{eqMinTmToGetFixedError} and $\tm$ decrease by the same percentage.

In the approximations used, the optimal parameters are the same for the case of a multi-photon input.
In the dispersive approximation, the system is linear with respect to the number of input photons.
Therefore, the contrast of the multi-photon readout depends only on the probability of a single photon passing through the cavity when the photon is off-resonant and the probability of reflecting it when the photon is in resonance with the cavity.
However, as explained for Eq.~\eqref{eqContrastApproximation}, these probabilities are proportional to $1-C$.
By increasing $C$ these probabilities decrease, which increases the contrast in the multi-photon case.

\begin{table}[t!]
	\caption{%
		Parameters for high-fidelity readout.
		Contrasts $C_\text{d}$ are calculated in the dispersive approximation using Eq.~\eqref{eqContrastApproximation}.
		Contrasts $C_\text{n}$ and the post-measurement qubit populations $P_\uparrow(\tm)$ are obtained numerically.
		An ideal detection is assumed with $\eta = 1$.
		The integration time relates to the pulse duration as $\tm = 7t_\ph$.
		For analytical estimates of parameters, $\tm = \TPurcell/15$ is chosen.
	}
	\begin{center}
		\label{tablHiFiReadout}
		\begin{ruledtabular}
			\newcolumntype{1}{D{.}{.}{1}} 
			\newcolumntype{2}{D{.}{.}{2}} 
			\newcolumntype{3}{D{.}{.}{3}} 
			\begin{tabular}{223111ccc}
				\hel{$\omega_\q/2\pi$} & \he{$\omega_\res/2\pi$}
				& \he{$\lambda$}
				& \he{$g/2\pi$}	& \he{$\kappa/2\pi$}
				& \he{$\tm$} & \he{$C_\text{d}$} & \he{$C_\text{n}$}
				& \he{$P_\uparrow$}
				\\
				\hel{(GHz)} & \he{(GHz)}
				& 
				& \he{(MHz)} & \he{(kHz)}
				& \he{(ms)}	& \he{(\%)} & \he{(\%)}
				& \he{(\%)}
				\\
				\hline
				\multicolumn{9}{l}{Parameters optimized analytically:}
				\\
				5.00 & 4.09
				& 0.006
				& 5.7 & 7.2
				& 36.9 & 99.3 & 98.1
				& 92.3
				\\
				20.00 & 16.36
				& 0.006
				& 22.9 & 28.9
				& 9.2 & 99.3 & 98.1
				& 92.2
				\\
				\hline
				\multicolumn{9}{l}{Parameters optimized numerically:}
				\\
				5.00 & 4.09
				& 0.005
				& 4.4 & 3.8
				& 36.9 & 98.8 & 98.5
				& 97.5 
				\\
				20.00 & 16.36
				& 0.005
				& 18.1 & 16.9
				& 9.2 & 98.8 & 98.5
				& 97.2 
			\end{tabular}
		\end{ruledtabular}
	\end{center}
\end{table}

\section{Effect of the Purcell decay}
\label{secRelaxation}

\begin{figure}[t!]
	\centering
	\includegraphics[width = 0.45\textwidth]{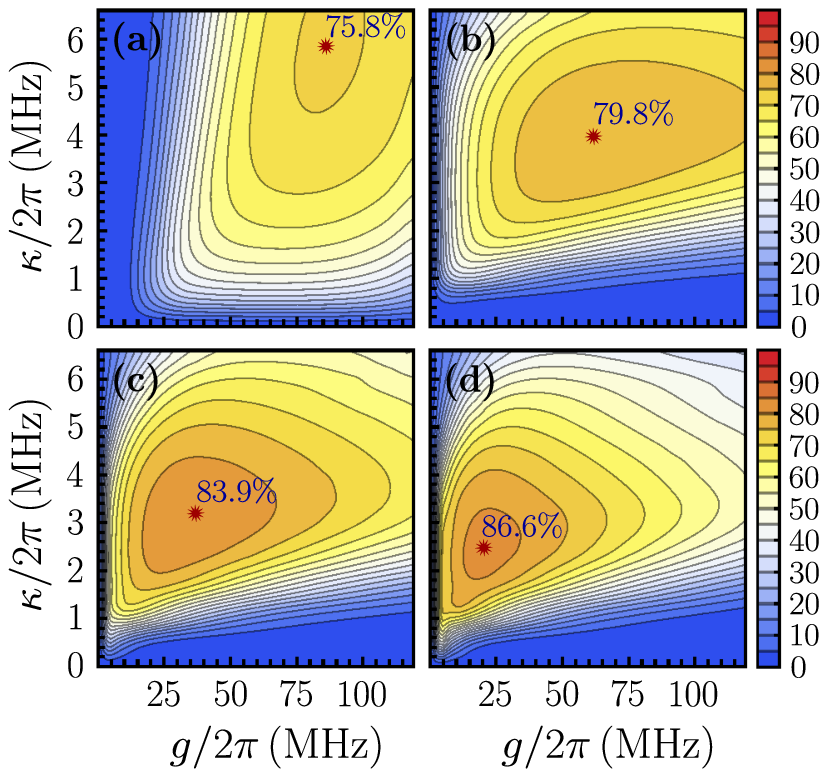}
	\caption{Dependence of the readout contrast on the resonator decay rate $\textstyle \kappa$ and the resonator-qubit coupling strength $\textstyle g$ for (a) $\textstyle \tm = 1\mu s$, (b) $\textstyle \tm = 2\mu s$, (c) $\textstyle \tm = 5\mu s$, and (d) $\textstyle \tm = 10\mu s$.
		Measurement time and pulse duration are related as $\textstyle \tm = 6 t_\ph$.
		The resonator and the qubit frequencies are $\textstyle \omegar = 4.09 \mathrm{GHz}$ and $\textstyle \omegaq=5.0 \mathrm{GHz}$, respectively.
		The photodetector is set to be ideal with $\textstyle \eta=1$.
		Star marks the position of the maximal contrast $\textstyle C_\mathrm{max}$ which can be achieved for given qubit-resonator detuning and readout pulse duration.}
	\label{figContr}
\end{figure}

In this section, we go beyond the dispersive approximation and study how the Purcell decay influences the readout performance.
This allows us to check the validity of the analytical results obtained in the previous sections.
Moreover, by taking the relaxation into account one can further enhance the contrast.
For this purpose, we use the Hamiltonian~\eqref{eqBSHamiltonian}.
Note that we do not account for the additional qubit relaxation which arises due to the effective qubit-waveguide coupling described by Eq.~\eqref{eqHwqBlochSiegert}.
This relaxation, however, is negligible compared to the Purcell decay due to the condition~\eqref{eqNegligibleNonRWARelaxation}.
Note the effect of Purcell decay on the readout with a coherent state and a photon number resolving detector was studied in Ref.~\cite{nesterov2019counting}.

The analytical model developed above relies on the fulfillment of the condition~\eqref{eqRelaxationNegligible}.
This condition implies that the qubit relaxation is the slowest process in the system.
Therefore, one can consider that the population of the qubit is constant during the readout, and the qubit-dependent shifted frequency of the resonator $\textstyle \mean{\omegaResEff}=\omega_\res + \chi\mean{\sigmaz}$ does not change in time as well.
The probe photon frequency is set to match the shifted frequency of the resonator.
Thus within the analytical approach the only origin of readout errors is the non-monochromaticity of the probe photon which results in its unwanted scattering.

First let us discuss the cases when the analytical approach works well.
The higher the post-readout qubit population $P_\uparrow$ is, the better $C_\text{d}$ approximates $C_\text{n}$.
This is seen from Tables~\ref{tablHiFiReadout} and~\ref{tablFastReadout}.
As for the parameters optimization, here the analytics provides good results if high fidelities are targeted.
We compare the resulting contrasts with those obtained by numerical optimization;
the details of the numerical method are described below.
One can see from Table~\ref{tablHiFiReadout} that numerical optimization gives only a slight improvement of $<$0.5\% for the contrast.

\begin{figure}[t!]
	\centering
	\includegraphics[width = 0.4\textwidth]{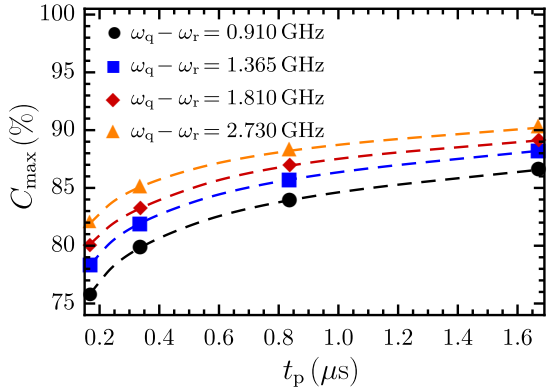}
	\caption{Dependence of the maximal readout contrast on the readout pulse duration $\textstyle t_\ph$ for different qubit-resonator detunings.
		To satisfy the criterion~\eqref{eqNegligibleNonRWARelaxation} we choose $\textstyle \lambda/\Lambda=10$.
		The rest of the parameters and notations are the same as in Fig.~\ref{figContr}.}
	\label{figContrOpt1}
\end{figure}

Note the readout times given in Table~\ref{tablHiFiReadout} are far beyond the best currently accessible lifetimes of superconducting qubits~\cite{rigetti2012superconducting,paik2016experimental,yan2016fluxrevisited}.
The contrasts given in the table do not take into account the intrinsic sources of the qubit decay, which start to play a role at such times.
The purpose of this table is to demonstrate the regime when the formula~\eqref{eqContrastApproximation} and the analytical optimization work perfectly well.

As follows from Eq.~\eqref{eqContrastApproximation}, to increase the readout contrast, one needs to decrease $\textstyle \kappa \tm$.
For this purpose, one can either extend the probe pulse duration $\textstyle t_p$ or increase the resonator decay rate $\textstyle \kappa$.
The use of a longer pulse slows down the readout, which limits the applicability of the scheme.
Either of these strategies can lead to a violation of the condition~\eqref{eqRelaxationNegligible} and break down the analytical approach.
In this case, the qubit relaxation comes into play, and its effect should be taken into account in the calculations.

Let us provide some general considerations regarding the effect of the qubit relaxation on the readout performance.
In the course of the readout, the qubit excited state decays, and the shifted frequency of the resonator drifts away from the probe photon frequency $\textstyle \omega_\ph$. This deteriorates the contrast.
To mitigate this issue and improve the contrast, one can suppress the qubit relaxation, i.e., increase $\textstyle \TPurcell = 1/(\lambda^2 \kappa)$.
To achieve this, one can reduce either $\textstyle \kappa$ or $\textstyle \lambda=g/(\omegaq-\omegar)$.
The latter can be accomplished by either reducing the qubit-resonator coupling strength $\textstyle g$ or increasing the detuning $\textstyle \omegaq-\omegar$. 
However, each of these approaches has side effects.
The decrease of the resonator decay rate $\textstyle \kappa$ elevates the error caused by the probe photon non-monochromaticity.
Weakening of the coupling $\textstyle g$ or increase of the detuning $\omega_\q - \omega_\res$ extends $\textstyle \TPurcell$ but lowers the total resonator frequency shift $\textstyle \chi=g(\lambda+\Lambda)$.
On the other hand, the larger the frequency shift $\textstyle \chi$ compared to the resonator linewidth $\textstyle \kappa$, the better one can resolve the $\textstyle \ket\downarrow$ from the $\textstyle \ket\uparrow$ qubit state.
The above considerations lead us to the conclusion that for a given measurement duration $\textstyle \tm$ and a qubit-resonator detuning $\textstyle \omegaq-\omegar$, there should exist a combination of $\textstyle \kappa$ and $\textstyle g$ when the readout error drops to its minimum while the contrast reaches the maximum.

\begin{figure}[t!]
	\centering
	\includegraphics[width = 0.4\textwidth]{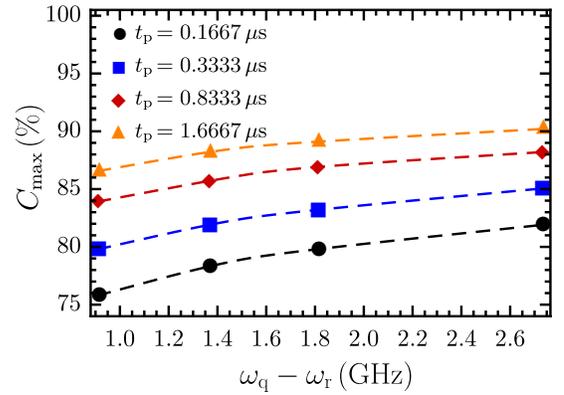}
	\caption{Dependence of the maximal readout contrast on the detuning between the qubit and resonator for different readout pulse durations $\textstyle t_\ph$.
		The rest of the parameters are as described in the caption of Fig.~\ref{figContrOpt1}.}
	\label{figContrOpt2}
\end{figure}

To illustrate this idea, we plot the readout contrast as a function of the qubit-resonator coupling $\textstyle g$ and the resonator decay rate $\textstyle \kappa$ for several measurement durations $\textstyle \tm = 6 t_\ph$ and a fixed value of the qubit-resonator detuning $\textstyle \omegaq-\omegar = 0.91 \, \mathrm{GHz}$.
To calculate the contrast, we use a theory that accounts for the Purcell decay of the qubit (see details in Appendix~\ref{apBeyondDispersive}).
The result is shown in Fig.~\ref{figContr}.
For each $\textstyle \tm$ we determine the maximal value of the contrast $\textstyle C_\mathrm{max}$ and the corresponding values of $\textstyle \kappa$ and $\textstyle g$.
Note that here we consider only the case $\textstyle \omegaq>\omegar$ due to the arguments presented in Section~\ref{secAnalytics}.

Using the method described above, we determine the dependence of $\textstyle C_\mathrm{max}$ on $\textstyle t_\ph$ and $\omega_\q - \omega_\res$.
Plots of these dependencies are shown in Figs.~\ref{figContrOpt1} and \ref{figContrOpt2}.
The obtained results demonstrate that a better contrast can be reached by using longer probe pulses and larger qubit-resonator detunings.
This conclusion agrees with Eq.~\eqref{eqErrorSetsLambda} derived within the analytical approach.

Numerical optimization provides considerably higher contrasts than analytical optimization if we optimize for a fast readout~(see Table~\ref{tablFastReadout}).
This is explained as follows.
As it was mentioned above, the qubit decay shifts the cavity resonance $\mean{\omegaResEff}$ during readout.
This raises the probability of unwanted scattering due to the detuning with the probe photon.
According to Eqs.~\eqref{eqEffectiveWr}, \eqref{eqTotalPull}, and~\eqref{eqOptCoupling}, the shift susceptibility to the decay increases proportionally to $\lambda$.
However, the pulse spectral density widens as $\lambda^5$ according to Eq.~\eqref{eqFinalTp}, which reduces unwanted scattering due to decay.
The reduction is more effective with larger $\lambda$.
Based on Eq.~\eqref{eqFinalTp}, it is the increase of $\lambda$ that is the best strategy to readout faster.
As explained before, one can sustain more qubit decay in that case.
Therefore, $\tm/\TPurcell$ should be increased to allow longer pulses and decrease the error~\eqref{eqErrorSetsLambda} due to the pulse non-monochromaticity.
This is not taken into account in the analytical optimization: see Eq.~\eqref{eqFinalTp}, where the ratio is fixed.
As the numerical optimization yields larger $\tm/\TPurcell$, the population $P_\uparrow$ is smaller for the relevant sets in Table~\ref{tablFastReadout}.
Note $P_\uparrow$ is better for the numerically optimized sets in Table~\ref{tablHiFiReadout}.
Indeed, $\lambda$ is chosen small there to achieve high contrasts.
However, $\tm/\TPurcell$ in analytical optimization is too large to get relaxation errors small enough.

\begin{table}[t!]
	\caption{%
		Parameters for fast readout.
		Here the measurement time is $\tm = 6t_\ph$, where $t_\ph$ is the pulse duration.
		For analytical parameter estimates we set $\tm = \TPurcell/10$.
		Other parameters, as well as notations, are the same as in Table~\ref{tablHiFiReadout}.
		To calculate $C_{\text d}$, Eqs.~\eqref{eqContrastDispersiveLorentzian}--\eqref{eqnDetuned} are used here as $\chi \tm$ does not satisfy the condition~\eqref{eqSimpleContrastCondition}.
	}
	\begin{center}
		\label{tablFastReadout}
		\begin{ruledtabular}
			\newcolumntype{1}{D{.}{.}{1}} 
			\newcolumntype{2}{D{.}{.}{2}} 
			\begin{tabular}{221121ccc}
				\hel{$\omega_\q/2\pi$} & \he{$\omega_\res/2\pi$}
				& \he{$\lambda$}
				& \he{$g/2\pi$}	& \he{$\kappa/2\pi$}
				& \he{$\tm$} & \he{$C_\text{d}$} & \he{$C_\text{n}$}
				& \he{$P_\uparrow$}
				\\
				\hel{(GHz)} & \he{(GHz)}
				& 
				& \he{(MHz)} & \he{(MHz)}
				& \he{($\mu$s)}	& \he{(\%)} & \he{(\%)}
				& \he{(\%)}
				\\
				\hline
				\multicolumn{9}{l}{Parameters optimized analytically:}
				\\
				5.00 & 4.09
				& 0.059
				& 53.6 & 4.08
				& 1.0 & 71.0 & 67.9
				& 89.2
				\\
				20.00 & 16.36 
				& 0.050
				& 180.0 & 9.99
				& 0.6 & 79.1 & 75.9
				& 88.9
				\\
				\hline
				\multicolumn{9}{l}{Parameters optimized numerically:}
				\\
				5.00 & 4.09
				& 0.095
				& 86.4 & 5.80
				& 1.0 & 82.0 & 75.8
				& 66.5 
				\\
				20.00 & 16.36 
				& 0.074
				& 269.0 & 12.90
				& 0.6 & 86.3 & 80.6
				& 71.9
			\end{tabular}
		\end{ruledtabular}
	\end{center}
\end{table}

\section{Discussion and outlook}
\label{secDiscussionOutlook}

A protocol for the dispersive readout that uses merely a single photon has been considered in the paper.
We have managed to develop an analytical model of the readout by neglecting the Purcell decay of the qubit.
Using this theory, we have derived a compact expression for the readout contrast.
Optimal parameters of the system have been expressed, too.
Both the readout time and its contrast are set by the characteristic frequencies of the system $\omega_\q$ and $\omega_\res$ and by the ratio $\lambda = g/(\omega_\q - \omega_\res)$ of the coupling strength to the detuning.
We have complemented our analytical approach with the numerical model, which accounts for the relaxation.
We have used the model to check the analytical contrasts and to optimize the system parameters further.
Making the measurement time closer to the qubit lifetime results in more relaxation, but gives less error due to the scattering.
Numerical optimization allows one to find a compromise between the relaxation and the scattering errors.
It is particularly helpful for designing a fast measurement reaching contrasts up to 90\%.
In that case, it gives an increase in contrast of more than 5\%.
For the contrasts above 98\%, numerics gives an improvement of about 0.5\%.
We stress that the only sources of errors in our scheme are qubit relaxation and unwanted scattering of the probe pulse.

There are no errors caused by the non-orthogonality of the states being distinguished by the detector.
Despite the absence of these errors, our scheme is slower than the state-of-the-art readout.
We attribute this to the fact that it uses just one photon.
As the photon is non-monochromatic, it can pass or reflect the cavity when it is not wanted.
By using more photons to probe the cavity, one can significantly decrease the probability of those errors.
For example, there are a few tens of photons in the measurement pulse in Ref.~\cite{walter2017rapid}, as can be shown by simple estimates with Eq.~(20) of Ref.~\cite{berman2012dynamics}.
The case of multiple input photons is left for future work.
Some of the results obtained in this paper might be helpful for the investigation of that case.
Due to the linearity of the system in the dispersive approximation, the scheme parameters optimized analytically are the same as for the single-photon case.
Besides, we have obtained the formula for the photon transport in the multi-photon case.
We also believe that the features of the system behavior due to the excitation exchange between the qubit and the resonator we have studied are qualitatively retained in the multi-photon case.

There are other possibilities to improve our scheme performance.
One can use stronger qubit-resonator coupling $g$ to obtain the higher magnitude $g(\lambda + \Lambda)$ of the qubit-dependent cavity pull.
To retain the non-demolition character of the readout, $\lambda=g/(\omega_\q - \omega_\res)$ and the Purcell decay rate $\TPurcell^{-1} = \kappa \lambda^2$ should be kept constant.
However, to minimize quasiparticle generation, the cavity and the qubit frequencies $\omega_\res$ and $\omega_\q$ are limited by the superconducting gap.
One can only increase the $|\Lambda/\lambda| = |\omega_\q - \omega_\res|/(\omega_\q + \omega_\res)$ ratio to overcome the Purcell decay.
While doing so additionally improves the readout due to the higher Bloch-Siegert shift, one would need to account for the qubit relaxation due to the counter-rotating terms in the Hamiltonian.
Interestingly, this type of relaxation depends on the combination of the qubit-resonator and the resonator-waveguide couplings~(see Ref.~\cite{bambaogawa2014recipe} and Appendix~\ref{apOtherCouplings}).
Alternatively, a Purcell filter~\cite{walter2017rapid} could be used to suppress the qubit relaxation while increasing the coupling $g$.
In this case, $\kappa_\q$ the resonator decay rate as seen by the qubit differs significantly from the resonance decay rate $\kappa$.
One can check that Eqs.~\eqref{eqMaxTpDueToQubitRelaxation}, \eqref{eqErrorSetsLambda}, and~\eqref{eqFinalTp} are then modified with the replacement $\textstyle t_\ph/\TPurcell \to (\kappa/\kappa_\q)(t_\ph/\TPurcell)$. 
Furthermore, if $\lambda$ is replaced with $\lambda \sqrt{\kappa/\kappa_\q}$, the measurement error $\varepsilon$~\eqref{eqErrorSetsLambda} does not change.
The measurement time $\tm \propto t_\ph$, however, decreases by the ratio of $\kappa/\kappa_\q$.
As can be deduced from Ref.~\cite{jeffrey2014fast}, $\kappa/\kappa_\q \sim 100$ is achievable for the typical parameters we use.

We expect that the studied setup for the superconducting qubit readout is favorable for on-chip integration.
On-chip circulators were already demonstrated~\cite{chapman2017widely,mahoney2017onchip,bernier2017nonreciprocal}.
The same holds for single-photon sources~\cite{peng2016tuneable,forndiaz2017on} and photodetectors~\cite{chen2011microwave,opremcak2018measurement,oelsner2018switching,inomata2016single,besse2018singleshot}.
Moreover, we only need one type of classical signals: those to prepare the states of the qubit and the photon source.
This may simplify the integration of control circuitry on a chip using the single flux quantum logic.
Such control was already demonstrated~\cite{kaplunenko2004experimental}, and some promising proposals for it were put forward~\cite{liebermann2016optimal,klenov2017flux,mcdermott2014accurate}.

In addition to the study of the single-photon readout, our work provides some general results.
We have treated a cavity quantum electrodynamics system with a method based on the Heisenberg-Langevin equations.
This approach has allowed us to make the RWA naturally for the coupling to the waveguides and to highlight the condition~\eqref{eqRWA} of its validity.
We have considered a situation when the RWA for the qubit-resonator coupling breaks, but the counter-rotating terms in the Hamiltonian can be treated as a perturbation.
For this case, a theory of multi-photon transport through the system was developed in the dispersive regime, neglecting the exchange of excitations between the resonator and the qubit.
A theory of single-photon transport through the system was developed while taking into account the excitation exchange.
We have shown that under the conditions~\eqref{eqBlochSiegert}, \eqref{eqWqWrSameOrderOfMagnitude}, and~\eqref{eqNegligibleNonRWARelaxation}, a change of coupling types does not change the magnitude of the dispersive pull.
Hence it does not alter the performance of a dispersive readout with any type of detector and probe.
Also, it has been found that the Bloch-Siegert shift can aid dispersive readout.
A proper choice of its sign increases the qubit-dependent cavity pull without a substantial impact on the qubit lifetime.

In conclusion, a theory of a single-photon dispersive measurement with a photodetector has been developed.
Using this theory, we have assessed the performance of the scheme.
Hence we were able to quantitatively analyze an ultimate limit of dispersive readout with an elementary portion of electromagnetic energy---a single photon.
Sources of the readout errors have been identified.
The role of the Bloch-Siegert shift and the coupling types has been elucidated.
Some of our results are also valid for the multi-photon input.
For future work, considering multi-photon Fock pulses is of interest.

\begin{acknowledgments}
Work of A.S.~was partly supported by DAAD grant~(2016).
The authors thank Frank Wilhelm-Mauch and Oleksandr Chumak for useful comments and for reading the manuscript.
A.S.~also acknowledges hospitality of Frank's group during the visits in 2016--2017.
We thank Oleksandr Chumak for suggesting the direction for this research.
\end{acknowledgments}

\appendix

\section{Other types of qubit-resonator and resonator-waveguide couplings}
\label{apOtherCouplings}

The coupling Hamiltonians~\eqref{eqHqr} and~\eqref{eqHwr} can be of more general form.
If any type of linear transversal coupling is allowed,
\begin{gather}
\label{eqHqrGeneral}
	H^{\text{gen}}_{\q\res} = \hbar (g^* \sigma_+ a + G \sigma_- a + \hc),
\\
\label{eqHwrGeneral}
	H^{\text{gen}}_{\res\alpha} = \hbar \int_0^\infty dk
		(f_k b_k^{\alpha\dag} a + F_k b_k a + \hc),
\end{gather}
where the constants of interaction $g$, $f_k$, and $F_k$ are complex.
Each coupling here is a mixture of inductive coupling, capacitive coupling, and a coupling described by charge-quasiflux terms like $Q\Phi$.

One partial case of Eqs.~\eqref{eqHqrGeneral} and~\eqref{eqHwrGeneral} is important from a practical standpoint.
In the main part of the paper, the qubit-resonator coupling is capacitive, and the resonator-waveguide coupling is inductive, or vice versa.
Let us consider the case when the qubit-resonator and the resonator-waveguide interactions are both either capacitive or inductive.
To describe it, one could alter the Hamiltonian~\eqref{eqHwr} of the cavity-waveguide coupling to
\begin{gather}
\label{eqHwrCapInd}
	H^{\text{same-type}}_{\res\alpha} = \hbar \int_0^\infty dk f_k
		(a + a^\dag)(b^\alpha_k + b_k^{\alpha\dag}).
\end{gather}

Should we use these Hamiltonians, our analytical treatment would change only trivially.
Consider the general case of Hamiltonians~\eqref{eqHqrGeneral} and~\eqref{eqHwrGeneral}.
It is straightforward to generalize the unitary transforms~\eqref{eqBlochSiegertTransform} and~\eqref{eqDispersiveTransform} to that case.
The total magnitude of the qubit-dependent shift of the cavity resonance changes to
\begin{equation}
	\chi = \Real(g^*\lambda + G^*\Lambda).
\end{equation}
In the case of the same-type couplings, $\chi$ stays intact and is given by Eq.~\eqref{eqTotalPull}.

For the numerical treatment, Hamiltonian in the Bloch-Siegert frame matters.
Again, consider the general case.
The effective qubit-waveguide Hamiltonians~\eqref{eqHwqBlochSiegert} are then given by
\begin{equation}
	H_{\q\alpha} = - \hbar \int_0^\infty dk
								(\Lambda^* F_k b^{\alpha\dag}_k \sigma_-
								+ \Lambda f^*_k b^\alpha_k \sigma_-).
\end{equation}
For example, in the same-type coupling case, the sign of $H_{\q\alpha}$~\eqref{eqHwqBlochSiegert} changes.
In any case, subsequent changes in the Heisenberg equations in Appendix~\ref{apBeyondDispersive} are of the order of $\Lambda^2$.
They are negligible under the condition~\eqref{eqNegligibleNonRWARelaxation}.

\section{Derivation of the approximate formula for the readout contrast}
\label{apApproximating}
Here we show how to approximate Eqs.~\eqref{eqContrastDispersiveLorentzian}--\eqref{eqnDetuned} with Eq.~\eqref{eqContrastApproximation}.
Equations~\eqref{eqContrastDispersiveLorentzian}--\eqref{eqnDetuned} yield
\begin{multline}
\label{eqContrastDispersiveFull}
	\frac C {\eta} = \frac K {K+1} -
				\frac{4K^2}{(K-1)^2 + 16X^2}
\\
					\times\left(\frac 1 4 + \frac 1 {4K} - \frac{K+1}
														{(K+1)^2 + 16X^2}\right).
\end{multline}
In what follows, we drop the terms that contribute below 0.1\% to $C/\eta$.
As $\eta \sim 1$, these terms contribute with the same order of magnitude to $C$ too.

The Taylor expansion of the first term in Eq.~\eqref{eqContrastDispersiveFull} is $K/(K+1) = 1 - 1/K + 1/2K^2 + \ldots$.
We drop the terms starting from $1/2K^2$, as they contribute to the order of magnitude $10^{-3}$ or less if $X \gtrsim 40$.
This can be shown using Eq.~\eqref{eqOptimalKForBigX}.
The second term in Eq.~\eqref{eqContrastDispersiveFull} is approximated as $K^2/16X^2$, which is below or of order $10^{-2}$ for $X \gtrsim 100$.
In the Taylor expansion $4K^2/[(K-1)^2 + 16X^2] = K^2/4X^2 [1 - (K-1)^2/16X^2 + \ldots]$ we neglect the terms starting with the second one.
$(K-1)^2/X^2 \sim 10^{-1}$ contributes as $10^{-3}$ to $C$ since $K \gtrsim 10$ for $X \gtrsim 100$.
Also, we neglected the terms $1/4K$ and $(K+1)/[(K+1)^2 + 16X^2]$ as these terms contribute not more than $10^{-3}$.
Finally,
\begin{equation}
	C \approx \eta \left(1 - \frac{1}{K} - \frac{K^2}{16X^2}\right) = \eta \left(1 - \frac{3}{2K}\right),
\end{equation}
for $X \gtrsim 100$; we used Eq.~\eqref{eqOptimalKForBigX} in obtaining the last equality.

\section{Theory of single-photon transport without the dispersive approximation}
\label{apBeyondDispersive}
In this Appendix, we complement the results presented in Section~\ref{secTransport} with a theory of a single-photon transport through the resonator-qubit system which does not rely on the dispersive approximation and accounts for the exchange of excitations between the qubit and the resonator.
This process ultimately leads to decay of the qubit excited state since the resonator is open (i.e., coupled to the transmission lines), which in turn affects the readout contrast.
Here we study the evolution of the system using the Hamiltonian $\textstyle H'$ expressed by Eq.~(\ref{eqBSHamiltonian}).
Since the criterion~\eqref{eqRWA} is fulfilled, we can apply the RWA in the Hamiltonian $\textstyle H'$ by neglecting the rapidly-oscillating terms $\textstyle \propto(b^{\alpha\dag}_k a^\dag + b^{\alpha}_k a)$ and $\textstyle \propto(b^{\alpha\dag}_k \sigma_+ + b^{\alpha}_k \sigma_-)$.
The contribution of these terms to the equations of motion is similar to that neglected in deriving Eqs.~\eqref{eqbkIntegrated}, \eqref{eqfkbkIntegrated}, and~\eqref{eqa}.
One can show that the Hamiltonian $\textstyle H'$ in the RWA conserves the number of excitations in the system:
\begin{equation} \label{eq:comm0}
 \left[H'_\mathrm{RWA}, N_\mathrm{ex}\right] = 0,
\end{equation}
where
\begin{equation}
 N_\mathrm{ex} = \int^\infty_0 dk \sum_{\alpha=\bi,\bii} b^{\alpha\dag}_k b^\alpha_k + a^\dag a + \sigma_+\sigma_-
\end{equation}
stands for the operator of the total number of excitations.

The Hamiltonian $\textstyle H'_\mathrm{RWA}$ generates the Heisenberg equations for the waveguide variables as follows:
\begin{equation} \label{eq:b_eqmot}
 \dot{b}^\alpha_k = - i \omega_k b^\alpha_k + f_k (a + \Lambda \sigma_-).
\end{equation}
Using the dispersion relation~(\ref{eqDispersion}), the formal solution of the above equation is written as
\begin{equation} \label{eq:bform}
 \begin{split}
  b^\alpha_k (t) = & \, b^\alpha_k(0)e^{-i v k t} \\
  & \, + f_k \int^t_0 dt' \, e^{-i v k (t-t')} (a + \Lambda \sigma_-)|_{t'}.
 \end{split}
\end{equation}
Equations of motion for the resonator and qubit variables read
\begin{equation} \label{eq:eqa_1}
  \dot{a} = - i (\omegar+g\Lambda\sigmaz) a - i g \sigma_- - \sum_{\alpha=\bi,\bii} \int^\infty_0 d k f_k b^\alpha_k
\end{equation}
and
\begin{equation} \label{eq:eqsigm_1}
 \begin{split}
  \dot{\sigma}_- = & \, - i \omegaq \sigma_- + i g \sigmaz a - i g\Lambda (2 a^\dag a + 1) \sigma_- \\
  & \, - \Lambda \sigmaz \sum_{\alpha=\bi,\bii} \int^\infty_0 d k f_k b^\alpha_k.
 \end{split}
\end{equation}
Using Eq.~(\ref{eq:bform}), one has
\begin{equation} \label{eq:intb1}
 \begin{split}
  \int^\infty_0 dk f_k b^\alpha_k(t) = & \, B^\alpha(t) + \int^\infty_0 dk f^2_k \int^t_0 dt' e^{-i v k (t-t')} a(t') \\
  & \, + \Lambda \int^\infty_0 dk f^2_k \int^t_0 dt' e^{-i v k (t-t')} \sigma_-(t'),
 \end{split}
\end{equation}
where we introduced a notation
\begin{equation}
 B^\alpha = \int^\infty_0 d k \, f_k b^\alpha_k(0) e^{-i v k t}.
\end{equation}
Following the similar considerations that led us from Eq.~\eqref{eqbkIntegratedRaw} to Eq.~\eqref{eqbkIntegrated}, we obtain
\begin{equation} \label{eq:intb}
  \sum_{\alpha=\bi,\bii} \int^\infty_0 dk f_k b^\alpha_k \approx \sum_{\alpha=\bi,\bii} B^\alpha + \frac{\kappa}{2} a + \Lambda \frac{\kappaq}{2} \sigma_-
\end{equation}
for $\textstyle t>0$.
It follows from Eq.~(\ref{eq:intb}) that the operator $\textstyle B^\alpha$ satisfies the commutation relations
\begin{equation} \label{eq:comms}
\left[B^\alpha, b^\alpha_k\right] = \left[B^\alpha, a\right] = \left[B^\alpha, \sigma_-\right] = 0.
\end{equation}
Substituting Eq.~(\ref{eq:intb}) into Eq.~(\ref{eq:eqa_1}), we obtain the equation
\begin{equation} \label{eq:a_eqmot}
  \dot{a} = - i \left(\bar{\omega}_\res + g\Lambda\sigmaz\right) a - i \gq \sigma_- - \sum_{\alpha=\bi,\bii} B^\alpha,
\end{equation}
where $\textstyle \bar{\omega}_\res = \omegar-i\kappa/2$ and $\textstyle \gq = g - i \Lambda \kappaq/2$.

Substituting Eq.~(\ref{eq:intb}) into Eq.~(\ref{eq:eqsigm_1}) and taking into account the condition~(\ref{eqNegligibleNonRWARelaxation}) gives
\begin{equation} \label{eq:sgm_eqmot}
  \dot{\sigma}_- = - i [\omega_\q + g\Lambda(2a^\dag a+1)]\sigma_- + i \gr \sigmaz a - \Lambda \sigmaz \sum_{\alpha=\bi,\bii} B^\alpha,
\end{equation}
where $\textstyle \gr = g + i \Lambda \kappa/2$.

\subsection{Readout contrast}
As we account for the excitation exchange between the qubit and the resonator, Eq.~\eqref{eqPclick} for the photodetection probability employed in the main part of the paper should be revisited here.
That expression holds provided that no more than one photon arrives at the detector.
However, when one prepares the qubit in the excited state, there is a non-zero probability to find two photons in the detector port (waveguide $\textstyle \bii$).
An extra photon can emerge due to the Purcell decay of the qubit.
In the general case of the multiphoton pulse arriving at the on--off photodetector, the probability of the detector click is determined as
$\textstyle \Pcl|_{q} = 1 - \left\langle : \exp(- \eta \Ntr) : \right\rangle_q$, where $\textstyle q=\uparrow, \downarrow$.
Expanding this expression in the Taylor series yields
\begin{equation}
\label{eqPcl}
 \Pcl|_{q} = \sum_{n\geq 1} (-1)^{n+1}\frac{\eta^n}{n!} \left\langle : \Ntr^n\!: \right\rangle_q,
\end{equation}
where $\textstyle : :$ denotes the normal ordering of operators, and the expectation value is calculated for the initial state $\textstyle |\Psi_q\rangle$.
The latter is given in the Bloch-Siegert picture, $\textstyle |\Psi_q\rangle = \UBS |\psi_q\rangle$ for the initial state in the dispersive frame $\textstyle \ket{\psi_q}$ expressed by Eq.~\eqref{eqInitialState}.

\subsubsection{Qubit prepared in the ground state}
When the qubit is prepared in the ground state, we face a single-excitation problem since the incident pulse contains only one photon, and the number of excitations in the system is conserved due to Eq.~\eqref{eq:comm0}.
In that case, the terms with $\textstyle n\ge 2$ vanish in Eq.~\eqref{eqPcl}.
Employing the identity  $\textstyle \Ntr = \int^\infty_0 dk b^{\bii\dag}_k b^\bii_k$ \cite{chumak2012operator, chumak2013phase}, one obtains
\begin{equation}
 \label{eqPcl_dn}
 \Pcl|_{\downarrow} = \eta \int^\infty_0 d k \, \langle\Psi_\downarrow|b^{\bii\dag}_k b^\bii_k \psidn.
\end{equation}
The initial state of the system in the Bloch-Siegert picture reads $\textstyle |\Psi_\downarrow\rangle \equiv \UBS |\psi_\downarrow\rangle= \ket\downarrow \ket{0^\res} \ket{1^\bi_\xi} \ket{0^\bii}$.

To calculate the integrand in Eq.~\eqref{eqPcl_dn}, we employ an approach similar to that presented in Ref. \cite{stolyarov2019few} (see Appendix A therein).
Using the representation $\textstyle b^{\bii}_k(t) = e^{i H' t} b^{\bii}_k(0) e^{-i H' t}$, one obtains
\begin{equation} \label{eqBb0}
 \langle\Psi_\downarrow|b^{\bii\dag}_k b^\bii_k \psidn = \langle \psi_1(t)|b^{\bii\dag}_k(0) b^\bii_k(0) \ket{\psi_1(t)},
\end{equation}
where $\textstyle \ket{\psi_1(t)} = e^{-i H' t}\psidn$ stands for the state of the system at the instant $\textstyle t$ provided that the initial state is $\textstyle \ket{\psi_1(0)} = \psidn$.
Since the number of excitations in the system is constant, its evolution occurs, in this case, only within the single-excitation domain of the Hilbert space of the system states.
The time-dependent single-excitation state $\textstyle \ket{\psi_1(t)}$ reads~\cite{stolyarov2019few}
\begin{equation} \label{eqpsidn}
\begin{split}
\ket{\psi_1(t)} = & \, \sum_{\alpha=\bi,\bii} \int^\infty_0 dk Z^\alpha_k(t)b^{\alpha\dag}_k(0)\ket{0} \\
& \, + Z_1(t) a^\dag(0)\ket{0} + Z_2(t) \sigma_+(0)\ket{0},
\end{split}
\end{equation}
where $\textstyle \ket 0 \equiv \ket \downarrow \ket{0^\mathrm{r}} \ket{0^\mathrm{I}} \ket{0^\mathrm{II}}$.
Substituting Eq.~\eqref{eqpsidn} on the rhs of Eq.~\eqref{eqBb0} results in
\begin{equation}
 \langle\Psi_\downarrow|b^{\bii\dag}_k b^\bii_k \psidn = |Z^\bii_k(t)|^2.
\end{equation}
It follows directly from Eq.~\eqref{eqpsidn} that $\textstyle Z^\bii_k(t) = \langle 0|b^\bii_k(0)|\psi_1(t)\rangle$, which gives $\textstyle Z^\bii_k(t) = e^{-i E_0 t} \langle 0|b^\bii_k(t)\psidn$, where $E_0 = \langle 0|H'|0\rangle$.
Finally, this leads to
\begin{equation} \label{eqBbdn}
 \langle\Psi_\downarrow|b^{\bii\dag}_k b^\bii_k \psidn = \left|\lvac b^\bii_k \psidn\right|^2.
\end{equation}

Using Eq.~(\ref{eq:b_eqmot}) along with the dispersion relation Eq.~(\ref{eqDispersion}), we derive the equation of motion for $\textstyle \lvac b^\bii_k \psidn$:
\begin{equation}
 \label{eqB01}
 \left(\partial_t + i v k\right) \lvac b^\bii_k \psidn = f_k \lvac a \psidn + \Lambda f_k \lvac\sigma_- \psidn,
\end{equation}
with the initial condition $\textstyle \lvac b^\bii_k \psidn = 0$ at $\textstyle t=0$.

Using Eqs.~(\ref{eq:a_eqmot}) and~(\ref{eq:sgm_eqmot}), one obtains the evolution equations for $\textstyle \lvac a\psidn$ and $\textstyle \lvac \sigma_-\psidn$, which read as
\begin{subequations}
 \label{eqA01S01}
 \begin{gather}
  \label{eqA01}
 \left[\partial_t + i (\bar{\omega}_\res - g\Lambda)\right]\lvac a \psidn = - i \gq \lvac \sigma_- \psidn - f_\ph \Xi(v t), \\
  \label{eqS01}
 \left[\partial_t + i (\omega_\q + g\Lambda)\right]\lvac \sigma_- \psidn = - i \gr \lvac a  \psidn - \Lambda f_\ph \Xi(vt).
 \end{gather}
\end{subequations}
The initial conditions for the above pair of equations are zero.
To derive Eqs.~\eqref{eqA01} and~\eqref{eqS01}, we accounted for $\textstyle \langle 0|a^\dag = 0$ and $\textstyle \langle 0|\sigmaz = -\langle 0|$.
Also, we employed $\textstyle B^\bii(t)|\Psi_\downarrow\rangle = 0$ and $\textstyle B^\bi(t)|\Psi_\downarrow\rangle \approx f_\ph \Xi(vt)|0\rangle$, where $\textstyle \Xi(v t) = \sqrt{2\pi} \Fourier[\xi'(k)](t)$ and $\textstyle f_\ph \equiv f_{k_\ph}$ with $\textstyle k_\ph$ being the central wave vector of an incident pulse.

The solutions of Eqs.~\eqref{eqB01} and~\eqref{eqA01S01} are written as
\begin{widetext}
 \begin{equation}
   \lvac b^\bii_k \psidn = i f_k f_\ph \int^t_0 dt' \, \Xi(v t') \left[\frac{\omegaq-g\Lambda-vk}{(\cE^+-vk)(\cE^--vk)}e^{-i v k (t-t')} + \sum_{\mu=\pm} \mu \frac{\omegaq-g\Lambda-\cE^\mu}{(\cE^+-\cE^-)(\cE^\mu - v k)} e^{-i\cE^\mu (t-t')}\right],
 \end{equation}
and
\begin{subequations}
 \begin{equation}
  \label{eqA01sln}
  \lvac a \psidn = \frac{f_\ph}{\cE^+-\cE^-} \int^t_0 dt' \, \Xi(v t')
  \sum_{\mu=\pm} \mu (\omegaq - \cE^\mu) e^{-i \cE^\mu (t-t')},
 \end{equation}
 \begin{equation}
  \label{eqS01sln}
  \lvac \sigma_- \psidn = \frac{f_\ph}{\cE^+-\cE^-} \int^t_0 dt' \, \Xi(v t') \sum_{\mu=\pm} \mu [\Lambda(\omegar - \cE^\mu) - g] e^{-i \cE^\mu (t-t')},
 \end{equation}
\end{subequations}
where
 \begin{equation*}
  \cE^\pm = \frac{\bar{\omega}_\res + \omega_\q}{2} \pm \sqrt{g_\res g_\q + \left(\frac{\bar{\omega}_\res - \omega_\q}{2} - g\Lambda\right)^2}
 \end{equation*}
are the single-photon resonances of the resonator-qubit system.
In Eqs.~\eqref{eqA01sln} and \eqref{eqS01sln} we have omitted terms $\textstyle \propto \Lambda^2$ due to the condition~\eqref{eqBlochSiegert}.
%
\subsubsection{Qubit prepared in the excited state}
When the qubit is prepared in the excited state, one deals with the two-excitation problem.
Then the probability of the detector click Eq.~\eqref{eqPcl} reduces to
\begin{equation}
\label{eqPcl_up}
   \Pcl|_{\uparrow} = \eta \int^\infty_0 d k \, \langle\Psi_\uparrow|b^{\bii\dag}_k b^\bii_k \psiup - \frac{\eta^2}{2} \int^\infty_0 d k \int^\infty_0 d k' \, \langle\Psi_\uparrow | b^{\bii\dag}_{k'} b^{\bii\dag}_k b^\bii_k b^\bii_{k'}|\Psi_\uparrow\rangle.
\end{equation}
The initial state in the Bloch-Siegert picture reads $\textstyle |\Psi_\uparrow\rangle = \ket\varphi \ket{1^\bi_\xi} \ket{0^\bii}$, where $\textstyle \ket\varphi \equiv \UBS|0^\res\rangle\ket\uparrow \approx \ket\uparrow \ket{0^\res} + \lambda \ket \downarrow \ket {1^\res}$.

To calculate the integrands in Eq.~\eqref{eqPcl_up}, we start with the time-dependent two-excitation state of the system $\textstyle \ket{\psi_2(t)} = e^{-i H' t}\psiup$.
The latter is expressed as \cite{stolyarov2019few}
\begin{equation} \label{eqpsiup}
\begin{split}
 \ket{\psi_2(t)} = & \, \sum_{\alpha=\bi,\bii} \sum_{\alpha'=\bi,\bii}\frac{1}{(2-\delta_{\alpha,\alpha'})\sqrt{1+\delta_{\alpha,\alpha'}}}\int^\infty_0 dk \int^\infty_0 dk' \varPhi^{\alpha,\alpha'}_{k,k'}(t) b^{\alpha\dag}_k(0) b^{\alpha'\dag}_{k'}(0)\ket{0} \\
  & \, + \sum_{\alpha=\bi,\bii} \int^\infty_0 dk \left[X^\alpha_k(t) a(0) + Y^\alpha_k(t) \sigma_-(0)\right]b^{\alpha\dag}_k(0)\ket{0} + \frac{1}{\sqrt{2}}Q_1(t) [a^\dag(0)]^2\ket{0} + Q_2(t) a^\dag(0)\sigma_+(0)\ket{0}.
\end{split}
\end{equation}
For the first integrand in Eq.~\eqref{eqPcl_up}, one has
\begin{equation} \label{eqBb2}
 \langle\Psi_\uparrow|b^{\bii\dag}_k b^\bii_k \psiup|_t = \langle\psi_2(t)|b^{\bii\dag}_k(0) b^\bii_k(0)\ket{\psi_2(t)} = 2 \int^\infty_0 dk'\left|\varPhi^{\bii,\bii}_{k,k'}(t)\right|^2 + \left|X^\bii_k(t)\right|^2 + \left|Y^\bii_k(t)\right|^2.
\end{equation}
For the second integrand, one obtains
\begin{equation} \label{eqBBbb2}
 \langle \Psi_\uparrow |b^{\bii\dag}_{k'} b^{\bii\dag}_k b^\bii_k b^\bii_{k'}|\Psi_\uparrow\rangle|_t = \langle \psi_2(t)|b^{\bii\dag}_{k'}(0) b^{\bii\dag}_k(0) b^\bii_k(0) b^\bii_{k'}(0)|\psi_2(t)\rangle
 = 2 \left|\varPhi^{\bii,\bii}_{k,k'}(t)\right|^2.
\end{equation}
The probability amplitudes arising in Eqs.~\eqref{eqBb2} and \eqref{eqBBbb2} can be expressed as \cite{stolyarov2019few}: $\textstyle \varPhi^{\bii,\bii}_{k,k'} = \langle 0|b^\bii_k b^\bii_{k'}\psiup/\sqrt{2}$, $\textstyle X^\bii_k = \langle 0|b^\bii_k a\psiup$, and $\textstyle Y^\bii_k = \langle 0|b^\bii_k \sigma_-\psiup$. This representation leads to the result as follows:
\begin{equation} \label{eqBbup}
  \langle\Psi_\uparrow|b^{\bii\dag}_k b^\bii_k \psiup = \int^\infty_0 dk'\left|\lvac b^\bii_{k'} b^\bii_k\psiup\right|^2 + \left|\lvac b^\bii_k a\psiup\right|^2 + \left|\lvac b^\bii_k \sigma_-\psiup\right|^2
\end{equation}
and
\begin{equation} \label{eqBBbb}
 \langle \Psi_\uparrow |b^{\bii\dag}_{k'} b^{\bii\dag}_k b^\bii_k b^\bii_{k'}|\Psi_\uparrow\rangle
 = \left|\lvac b^\bii_k b^\bii_{k'} \psiup\right|^2.
\end{equation}

Employing Eq.~(\ref{eq:b_eqmot}), we obtain the equation of motion for $\textstyle \lvac b^\bii_k b^\bii_{k'} \psiup$:
\begin{equation} \label{eq:eq_bb}
	\left[\partial_t + i v (k + k')\right] \lvac b^\bii_k b^\bii_{k'} \psiup = \lvac (f_{k'} b^\bii_k + f_k b^\bii_{k'})(a +\Lambda\sigma_-)\psiup.
\end{equation}
Using Eqs.~(\ref{eq:b_eqmot}), (\ref{eq:a_eqmot}) and~(\ref{eq:sgm_eqmot}) along with the property~(\ref{eq:comms}), one derives the equations of motion for $\textstyle \lvac b^\bii_k a \psiup$ and $\textstyle \lvac b^\bii_k \sigma_-\psiup$ as follows:
 \begin{gather}
  \label{eq:eq_ba}
   \left[\partial_t + i (v k + \bar{\omega}_\res - g\Lambda)\right] \lvac b^\bii_k a \psiup = - i \gq \lvac b^\bii_k \sigma_- \psiup + f_k \lvac a^2\psiup + \Lambda f_k \lvac \sigma_- a \psiup - f_\ph \Xi(vt) \lvac b^\bii_k|\varPhi\rangle, \\
  \label{eq:eq_bs}
   \left[\partial_t + i(v k + \omega_\q + g\Lambda)\right] \lvac b^\bii_k \sigma_-\psiup = - i \gr \lvac b^\bii_k a\psiup + f_k \lvac \sigma_- a\psiup  - \Lambda f_\ph  \Xi(vt) \lvac b^\bii_k|\varPhi\rangle,
 \end{gather}
where $\textstyle \ket{\varPhi} = \ket{\varphi}\ket{0^\mathrm{I}}\ket{0^\mathrm{II}}$.
Analogously, one has
 \begin{gather}
  \label{eq:eq_a2}
  \left[\partial_t + 2i(\bar{\omega}_\res-g\Lambda)\right] \lvac a^2 \psiup = - i \gq \lvac \sigma_- a \psiup - 2 f_\ph \Xi(vt) \lvac a|\varPhi\rangle, \\
  \label{eq:eq_as}
  \left[\partial_t + i(\omega_\q+\bar{\omega}_\res)\right]\lvac \sigma_- a\psiup = - i \gr \lvac a^2\psiup - f_\ph \Xi(vt)\lvac (\sigma_- + \Lambda a)|\varPhi\rangle.
 \end{gather}
The initial conditions for Eqs.~(\ref{eq:eq_bb})--(\ref{eq:eq_as}) are zero.

Equations of motion for $\textstyle \lvac b^\bii_k|\varPhi\rangle$, $\textstyle \lvac a |\varPhi\rangle$, and $\textstyle \lvac \sigma_- |\varPhi\rangle$ read
\begin{subequations}
 \label{eqBAS0phi}
 \begin{gather}
  \left(\partial_t + i v k \right) \lvac b^\bii_k|\varPhi\rangle = f_k \lvac a |\varPhi\rangle + \Lambda f_k \lvac \sigma_-|\varPhi\rangle, \\
  \left[\partial_t + i (\bar{\omega}_\res-g\Lambda)\right] \lvac a |\varPhi\rangle = - i \gq \lvac \sigma_-|\varPhi\rangle, \\
  \left[\partial_t + i (\omega_\q+g\Lambda)\right] \lvac \sigma_- |\varPhi\rangle = - i \gr \lvac a|\varPhi\rangle.
 \end{gather}
\end{subequations}
The initial conditions (at $\textstyle t=0$) for the above set of equations are $\textstyle \lvac b^\bii_k|\varPhi\rangle = 0$, $\textstyle \lvac a |\varPhi\rangle = \lambda$, and $\textstyle \lvac \sigma_-|\varPhi\rangle = 1$.
Neglecting the terms $\textstyle \propto \lambda \Lambda$ due to the conditions~\eqref{eqBlochSiegert} and~\eqref{eqNegligibleNonRWARelaxation}, the solution of Eqs.~\eqref{eqBAS0phi} is given by
\begin{subequations}
 \begin{gather}
   \label{eq:eq_bphi}
   \lvac b^\bii_k |\varPhi\rangle = i f_k \left[\frac{g_\q + \lambda(\omega_\q - v k)}{(\cE^+-vk)(\cE^--vk)}e^{-i v k t}
   + \sum_{\mu=\pm} \mu \frac{g_\q - \lambda (\omega_\q - \cE^\mu)}{(\cE^+-\cE^-)(\cE^\mu - v k)} e^{-i\cE^\mu t}\right], \\
   \lvac a |\varPhi\rangle = \frac{1}{\cE^+-\cE^-} \sum_{\mu=\pm} \mu \left[g_\q-\lambda(\omega_\q - \cE^\mu)\right] e^{-i \cE^\mu t}, \\
   \lvac \sigma_- |\varPhi\rangle = - \frac{1}{\cE^+-\cE^-} \sum_{\mu=\pm} \mu \left[\bar\omega_\res-g(\lambda+\Lambda) - \cE^\mu\right] e^{-i \cE^\mu t}.
 \end{gather}
\end{subequations}
\end{widetext}

\subsection{Qubit population dynamics}
The population of the qubit is determined as $\textstyle P_q(t) = \at{\bra{\Psi_q}\sigma_+\sigma_-\ket{\Psi_q}}{t}$.
Following the lines of the derivation of Eqs.~\eqref{eqBbdn} and \eqref{eqBb2}, one obtains
\begin{equation} \label{eqPdn}
   P_\downarrow(t) = |Z_2(t)|^2 = |\bra 0 \sigma_-\ket{\Psi_\downarrow}|^2
\end{equation}
for the qubit prepared in the ground state and
\begin{equation} \label{eqPup}
 \begin{split}
  P_\uparrow(t) = & \, |Q_2(t)|^2 + \sum_{\alpha=\bi,\bii} \int^\infty_0 dk |Y^\alpha_k(t)|^2 \\
  = & \, \left|\bra 0 \sigma_- a\ket{\Psi_\uparrow}\right|^2 + \sum_{\alpha=\bi,\bii} \int^\infty_0 dk \left|\bra 0 b^\alpha_k \sigma_- \ket{\Psi_\uparrow}\right|^2
 \end{split}
\end{equation}
for the qubit prepared in the excited state.

The last term in the second line of Eq.~\eqref{eqPup} suggests that we need to determine the matrix element $\textstyle \bra 0 b^\bi_k \sigma_- \ket{\Psi_\uparrow}$.
The evolution equation for the latter is obtained by the replacement of waveguide indices from $\textstyle \bii$ to $\textstyle \bi$ in Eq.~\eqref{eq:eq_bs}.
Equations of motion for $\textstyle \bra 0 b^\bi_k a \ket{\Psi_\uparrow}$ and $\textstyle \lvac b^\bi_k |\varPhi\rangle$ are derived by the analogous replacement in Eqs.~\eqref{eq:eq_ba} and~\eqref{eq:eq_bphi}, correspondingly.
The initial conditions are $\textstyle \at{\bra 0 b^\bi_k \sigma_- \ket{\Psi_\uparrow}}{t=0} = \xi'(k)$ and $\textstyle \at{\bra 0 b^\bi_k a \ket{\Psi_\uparrow}}{t=0} = \lambda \xi'(k)$.
For $\textstyle \lvac b^\bi_k |\varPhi\rangle$, we have $\textstyle \at{\lvac b^\bi_k |\varPhi\rangle}{t=0} = \at{\lvac b^\bii_k |\varPhi\rangle}{t=0} = 0$ implying that $\textstyle \lvac b^\bi_k |\varPhi\rangle = \lvac b^\bii_k |\varPhi\rangle$.

\subsection{Computational details}
The system of equations~(\ref{eq:eq_bb})--(\ref{eq:eq_as}) is solved analytically via the Laplace transform.
For this task, we use \texttt{LaplaceTransform} and \texttt{InverseLaplaceTransform} functions of the \textit{Mathematica} system.
This approach substantially reduces the computation time to obtain the data from Sec.~\ref{secRelaxation}.
However, the derived expressions are cumbersome, so we do not present them in the paper.
The integrals over the wave vectors in Eqs.~\eqref{eqPcl_dn}, \eqref{eqPcl_up} and~\eqref{eqPup} are computed numerically employing \texttt{NIntegrate} routine of \textit{Mathematica}.

Let us briefly discuss how the computational cost of the wave function approach described in this Appendix changes for a multiphoton Fock input.
To determine the state of the system for the case of an $N$-photon probe pulse, one needs to solve a set of $(N+1)^2$ and $(N+2)^2$ differential equations governing the probability amplitudes for the qubit prepared in the ground and excited state, respectively.
The analytical solution becomes intractable already for $N\geq 2$.
Moreover, for computation of the detector click probabilities [see Eq.~\eqref{eqPcl}], it is required to perform $(N+1)$-fold integrations over the wave vectors.
Thus the problem rapidly becomes computationally demanding with the increase of the number of photons in the probe pulse.

\bibliography{common_sources,jj_sources,cqed,measurement,counters,array,circulators,few_photon_sources,sfq,amp}
\bibliographystyle{apsrev4-1}
\end{document}

%% file: poznachennya.tex
\usepackage{mathtools}

\newcommand{\res}{\mathrm{r}}
\newcommand{\q}{\mathrm{q}}
\newcommand{\ph}{\mathrm{p}}
\newcommand{\w}{\mathrm{w}}

\newcommand{\tm}{{t_\mathrm m}}
\newcommand{\taum}{\tau_{\mathrm m}}
\newcommand{\Purcell}{\mathrm P}
\newcommand{\TPurcell}{T_\Purcell}

\newcommand{\trans}{\mathrm{tr}}

\newcommand{\sigmax}{\sigma_x}
\newcommand{\sigmay}{\sigma_y}
\newcommand{\sigmaz}{\sigma_z}

\newcommand{\bi}{\mathrm{I}}
\newcommand{\bii}{\mathrm{II}}

\newcommand{\UBS}{U_\mathrm{BS}}
\newcommand{\Udisp}{U_\mathrm{d}}

\newcommand{\BS}[1]{#1'}
\newcommand{\BSd}[1]{\widetilde #1}
\newcommand{\omegaResEff}{\BSd\omega_\res}

\newcommand{\click}{\mathrm{cl}}

\newcommand{\Kernel}{R}

\newcommand{\omegar}{\omega_\res}
\newcommand{\omegaq}{\omega_\q}

\newcommand{\kappaq}{\kappa_\q}
\newcommand{\gr}{g_\res}
\newcommand{\gq}{g_\q}
\newcommand{\Pcl}{P_\click}
\newcommand{\lvac}{\langle 0|}

\newcommand{\psidn}{|\Psi_\downarrow\rangle}
\newcommand{\psiup}{|\Psi_\uparrow\rangle}
\newcommand{\Ntr}{\mathcal{N}_\mathrm{tr}}
\newcommand{\cE}{\mathcal{E}}

%% file: fizychni_komandy.tex
\usepackage{amsmath}

\newcommand{\hc}{\text{H.~c.}}
\newcommand{\cc}{\text{c.~c.}}

\newcommand{\mean}[1]{\langle{#1}\rangle}
\newcommand{\Bigmean}[1]{\Big<{#1}\Big>}

\newcommand{\bra}[1]{\langle{#1}|} 
\newcommand{\ket}[1]{|{#1}\rangle} 
\newcommand{\braket}[2]{\langle{#1}|{#2}\rangle} 



%% file: dlya_tablychok.tex
\usepackage{dcolumn}

\newcommand\he[1]{\multicolumn{1}c{#1}} 

\newcommand\hel[1]{\multicolumn{1}l{#1}} 

%% file: paper.bbl
\begin{thebibliography}{50}%
\makeatletter
\providecommand \@ifxundefined [1]{%
 \@ifx{#1\undefined}
}%
\providecommand \@ifnum [1]{%
 \ifnum #1\expandafter \@firstoftwo
 \else \expandafter \@secondoftwo
 \fi
}%
\providecommand \@ifx [1]{%
 \ifx #1\expandafter \@firstoftwo
 \else \expandafter \@secondoftwo
 \fi
}%
\providecommand \natexlab [1]{#1}%
\providecommand \enquote  [1]{``#1''}%
\providecommand \bibnamefont  [1]{#1}%
\providecommand \bibfnamefont [1]{#1}%
\providecommand \citenamefont [1]{#1}%
\providecommand \href@noop [0]{\@secondoftwo}%
\providecommand \href [0]{\begingroup \@sanitize@url \@href}%
\providecommand \@href[1]{\@@startlink{#1}\@@href}%
\providecommand \@@href[1]{\endgroup#1\@@endlink}%
\providecommand \@sanitize@url [0]{\catcode `\\12\catcode `\$12\catcode
  `\&12\catcode `\#12\catcode `\^12\catcode `\_12\catcode `\%12\relax}%
\providecommand \@@startlink[1]{}%
\providecommand \@@endlink[0]{}%
\providecommand \url  [0]{\begingroup\@sanitize@url \@url }%
\providecommand \@url [1]{\endgroup\@href {#1}{\urlprefix }}%
\providecommand \urlprefix  [0]{URL }%
\providecommand \Eprint [0]{\href }%
\providecommand \doibase [0]{http://dx.doi.org/}%
\providecommand \selectlanguage [0]{\@gobble}%
\providecommand \bibinfo  [0]{\@secondoftwo}%
\providecommand \bibfield  [0]{\@secondoftwo}%
\providecommand \translation [1]{[#1]}%
\providecommand \BibitemOpen [0]{}%
\providecommand \bibitemStop [0]{}%
\providecommand \bibitemNoStop [0]{.\EOS\space}%
\providecommand \EOS [0]{\spacefactor3000\relax}%
\providecommand \BibitemShut  [1]{\csname bibitem#1\endcsname}%
\let\auto@bib@innerbib\@empty
\bibitem [{\citenamefont {Blais}\ \emph {et~al.}(2004)\citenamefont {Blais},
  \citenamefont {Huang}, \citenamefont {Wallraff}, \citenamefont {Girvin},\
  and\ \citenamefont {Schoelkopf}}]{blais2004cavity}%
  \BibitemOpen
  \bibfield  {author} {\bibinfo {author} {\bibfnamefont {A.}~\bibnamefont
  {Blais}}, \bibinfo {author} {\bibfnamefont {R.-S.}\ \bibnamefont {Huang}},
  \bibinfo {author} {\bibfnamefont {A.}~\bibnamefont {Wallraff}}, \bibinfo
  {author} {\bibfnamefont {S.~M.}\ \bibnamefont {Girvin}}, \ and\ \bibinfo
  {author} {\bibfnamefont {R.~J.}\ \bibnamefont {Schoelkopf}},\ }\href
  {\doibase 10.1103/PhysRevA.69.062320} {\bibfield  {journal} {\bibinfo
  {journal} {Phys. Rev. A}\ }\textbf {\bibinfo {volume} {69}},\ \bibinfo
  {pages} {062320} (\bibinfo {year} {2004})}\BibitemShut {NoStop}%
\bibitem [{\citenamefont {Walter}\ \emph {et~al.}(2017)\citenamefont {Walter},
  \citenamefont {Kurpiers}, \citenamefont {Gasparinetti}, \citenamefont
  {Magnard}, \citenamefont {Poto\ifmmode~\check{c}\else \v{c}\fi{}nik},
  \citenamefont {Salath\'e}, \citenamefont {Pechal}, \citenamefont {Mondal},
  \citenamefont {Oppliger}, \citenamefont {Eichler},\ and\ \citenamefont
  {Wallraff}}]{walter2017rapid}%
  \BibitemOpen
  \bibfield  {author} {\bibinfo {author} {\bibfnamefont {T.}~\bibnamefont
  {Walter}}, \bibinfo {author} {\bibfnamefont {P.}~\bibnamefont {Kurpiers}},
  \bibinfo {author} {\bibfnamefont {S.}~\bibnamefont {Gasparinetti}}, \bibinfo
  {author} {\bibfnamefont {P.}~\bibnamefont {Magnard}}, \bibinfo {author}
  {\bibfnamefont {A.}~\bibnamefont {Poto\ifmmode~\check{c}\else
  \v{c}\fi{}nik}}, \bibinfo {author} {\bibfnamefont {Y.}~\bibnamefont
  {Salath\'e}}, \bibinfo {author} {\bibfnamefont {M.}~\bibnamefont {Pechal}},
  \bibinfo {author} {\bibfnamefont {M.}~\bibnamefont {Mondal}}, \bibinfo
  {author} {\bibfnamefont {M.}~\bibnamefont {Oppliger}}, \bibinfo {author}
  {\bibfnamefont {C.}~\bibnamefont {Eichler}}, \ and\ \bibinfo {author}
  {\bibfnamefont {A.}~\bibnamefont {Wallraff}},\ }\href {\doibase
  10.1103/PhysRevApplied.7.054020} {\bibfield  {journal} {\bibinfo  {journal}
  {Phys. Rev. Applied}\ }\textbf {\bibinfo {volume} {7}},\ \bibinfo {pages}
  {054020} (\bibinfo {year} {2017})}\BibitemShut {NoStop}%
\bibitem [{\citenamefont {Clarke}\ and\ \citenamefont
  {Wilhelm}(2008)}]{clarke2008superconducting}%
  \BibitemOpen
  \bibfield  {author} {\bibinfo {author} {\bibfnamefont {J.}~\bibnamefont
  {Clarke}}\ and\ \bibinfo {author} {\bibfnamefont {F.~K.}\ \bibnamefont
  {Wilhelm}},\ }\href {\doibase 10.1038/nature07128} {\bibfield  {journal}
  {\bibinfo  {journal} {Nature}\ }\textbf {\bibinfo {volume} {453}},\ \bibinfo
  {pages} {1031} (\bibinfo {year} {2008})}\BibitemShut {NoStop}%
\bibitem [{\citenamefont {Berman}\ \emph {et~al.}(2009)\citenamefont {Berman},
  \citenamefont {Bishop}, \citenamefont {Chumak}, \citenamefont {Kinion},\ and\
  \citenamefont {Tsifrinovich}}]{berman2009measurement}%
  \BibitemOpen
  \bibfield  {author} {\bibinfo {author} {\bibfnamefont {G.~P.}\ \bibnamefont
  {Berman}}, \bibinfo {author} {\bibfnamefont {A.~R.}\ \bibnamefont {Bishop}},
  \bibinfo {author} {\bibfnamefont {A.~A.}\ \bibnamefont {Chumak}}, \bibinfo
  {author} {\bibfnamefont {D.}~\bibnamefont {Kinion}}, \ and\ \bibinfo {author}
  {\bibfnamefont {V.~I.}\ \bibnamefont {Tsifrinovich}},\ }\href
  {https://arxiv.org/abs/0912.3791} {\bibfield  {journal} {\bibinfo  {journal}
  {arXiv:0912.3791}\ } (\bibinfo {year} {2009})}\BibitemShut {NoStop}%
\bibitem [{\citenamefont {Roy}\ and\ \citenamefont
  {Devoret}(2018)}]{roy2018quantum}%
  \BibitemOpen
  \bibfield  {author} {\bibinfo {author} {\bibfnamefont {A.}~\bibnamefont
  {Roy}}\ and\ \bibinfo {author} {\bibfnamefont {M.}~\bibnamefont {Devoret}},\
  }\href {\doibase 10.1103/PhysRevB.98.045405} {\bibfield  {journal} {\bibinfo
  {journal} {Phys. Rev. B}\ }\textbf {\bibinfo {volume} {98}},\ \bibinfo
  {pages} {045405} (\bibinfo {year} {2018})}\BibitemShut {NoStop}%
\bibitem [{\citenamefont {Govia}\ \emph {et~al.}(2014)\citenamefont {Govia},
  \citenamefont {Pritchett}, \citenamefont {Xu}, \citenamefont {Plourde},
  \citenamefont {Vavilov}, \citenamefont {Wilhelm},\ and\ \citenamefont
  {McDermott}}]{govia2014high}%
  \BibitemOpen
  \bibfield  {author} {\bibinfo {author} {\bibfnamefont {L.~C.~G.}\
  \bibnamefont {Govia}}, \bibinfo {author} {\bibfnamefont {E.~J.}\ \bibnamefont
  {Pritchett}}, \bibinfo {author} {\bibfnamefont {C.}~\bibnamefont {Xu}},
  \bibinfo {author} {\bibfnamefont {B.~L.~T.}\ \bibnamefont {Plourde}},
  \bibinfo {author} {\bibfnamefont {M.~G.}\ \bibnamefont {Vavilov}}, \bibinfo
  {author} {\bibfnamefont {F.~K.}\ \bibnamefont {Wilhelm}}, \ and\ \bibinfo
  {author} {\bibfnamefont {R.}~\bibnamefont {McDermott}},\ }\href {\doibase
  10.1103/PhysRevA.90.062307} {\bibfield  {journal} {\bibinfo  {journal}
  {Physical Review A}\ }\textbf {\bibinfo {volume} {90}},\ \bibinfo {pages}
  {062307} (\bibinfo {year} {2014})}\BibitemShut {NoStop}%
\bibitem [{\citenamefont {Sokolov}(2016)}]{sokolov2016optimal}%
  \BibitemOpen
  \bibfield  {author} {\bibinfo {author} {\bibfnamefont {A.}~\bibnamefont
  {Sokolov}},\ }\href {\doibase 10.1103/PhysRevA.93.032323} {\bibfield
  {journal} {\bibinfo  {journal} {Phys. Rev. A}\ }\textbf {\bibinfo {volume}
  {93}},\ \bibinfo {pages} {032323} (\bibinfo {year} {2016})}\BibitemShut
  {NoStop}%
\bibitem [{\citenamefont {Opremcak}\ \emph {et~al.}(2018)\citenamefont
  {Opremcak}, \citenamefont {Pechenezhskiy}, \citenamefont {Howington},
  \citenamefont {Christensen}, \citenamefont {Beck}, \citenamefont {Leonard},
  \citenamefont {Suttle}, \citenamefont {Wilen}, \citenamefont {Nesterov},
  \citenamefont {Ribeill}, \citenamefont {Thorbeck}, \citenamefont {Schlenker},
  \citenamefont {Vavilov}, \citenamefont {Plourde},\ and\ \citenamefont
  {McDermott}}]{opremcak2018measurement}%
  \BibitemOpen
  \bibfield  {author} {\bibinfo {author} {\bibfnamefont {A.}~\bibnamefont
  {Opremcak}}, \bibinfo {author} {\bibfnamefont {I.~V.}\ \bibnamefont
  {Pechenezhskiy}}, \bibinfo {author} {\bibfnamefont {C.}~\bibnamefont
  {Howington}}, \bibinfo {author} {\bibfnamefont {B.~G.}\ \bibnamefont
  {Christensen}}, \bibinfo {author} {\bibfnamefont {M.~A.}\ \bibnamefont
  {Beck}}, \bibinfo {author} {\bibfnamefont {E.}~\bibnamefont {Leonard}},
  \bibinfo {author} {\bibfnamefont {J.}~\bibnamefont {Suttle}}, \bibinfo
  {author} {\bibfnamefont {C.}~\bibnamefont {Wilen}}, \bibinfo {author}
  {\bibfnamefont {K.~N.}\ \bibnamefont {Nesterov}}, \bibinfo {author}
  {\bibfnamefont {G.~J.}\ \bibnamefont {Ribeill}}, \bibinfo {author}
  {\bibfnamefont {T.}~\bibnamefont {Thorbeck}}, \bibinfo {author}
  {\bibfnamefont {F.}~\bibnamefont {Schlenker}}, \bibinfo {author}
  {\bibfnamefont {M.~G.}\ \bibnamefont {Vavilov}}, \bibinfo {author}
  {\bibfnamefont {B.~L.~T.}\ \bibnamefont {Plourde}}, \ and\ \bibinfo {author}
  {\bibfnamefont {R.}~\bibnamefont {McDermott}},\ }\href {\doibase
  10.1126/science.aat4625} {\bibfield  {journal} {\bibinfo  {journal}
  {Science}\ }\textbf {\bibinfo {volume} {361}},\ \bibinfo {pages} {1239}
  (\bibinfo {year} {2018})}\BibitemShut {NoStop}%
\bibitem [{\citenamefont {Barzanjeh}\ \emph {et~al.}(2014)\citenamefont
  {Barzanjeh}, \citenamefont {DiVincenzo},\ and\ \citenamefont
  {Terhal}}]{divincenzo2014dispersive}%
  \BibitemOpen
  \bibfield  {author} {\bibinfo {author} {\bibfnamefont {S.}~\bibnamefont
  {Barzanjeh}}, \bibinfo {author} {\bibfnamefont {D.~P.}\ \bibnamefont
  {DiVincenzo}}, \ and\ \bibinfo {author} {\bibfnamefont {B.~M.}\ \bibnamefont
  {Terhal}},\ }\href {\doibase 10.1103/PhysRevB.90.134515} {\bibfield
  {journal} {\bibinfo  {journal} {Phys. Rev. B}\ }\textbf {\bibinfo {volume}
  {90}},\ \bibinfo {pages} {134515} (\bibinfo {year} {2014})}\BibitemShut
  {NoStop}%
\bibitem [{\citenamefont {Didier}\ \emph {et~al.}(2015)\citenamefont {Didier},
  \citenamefont {Kamal}, \citenamefont {Oliver}, \citenamefont {Blais},\ and\
  \citenamefont {Clerk}}]{didier2015heisenberg}%
  \BibitemOpen
  \bibfield  {author} {\bibinfo {author} {\bibfnamefont {N.}~\bibnamefont
  {Didier}}, \bibinfo {author} {\bibfnamefont {A.}~\bibnamefont {Kamal}},
  \bibinfo {author} {\bibfnamefont {W.~D.}\ \bibnamefont {Oliver}}, \bibinfo
  {author} {\bibfnamefont {A.}~\bibnamefont {Blais}}, \ and\ \bibinfo {author}
  {\bibfnamefont {A.~A.}\ \bibnamefont {Clerk}},\ }\href {\doibase
  10.1103/PhysRevLett.115.093604} {\bibfield  {journal} {\bibinfo  {journal}
  {Phys. Rev. Lett.}\ }\textbf {\bibinfo {volume} {115}},\ \bibinfo {pages}
  {093604} (\bibinfo {year} {2015})}\BibitemShut {NoStop}%
\bibitem [{\citenamefont {Peng}\ \emph {et~al.}(2016)\citenamefont {Peng},
  \citenamefont {de~Graaf}, \citenamefont {Tsai},\ and\ \citenamefont
  {Astafiev}}]{peng2016tuneable}%
  \BibitemOpen
  \bibfield  {author} {\bibinfo {author} {\bibfnamefont {Z.~H.}\ \bibnamefont
  {Peng}}, \bibinfo {author} {\bibfnamefont {S.~E.}\ \bibnamefont {de~Graaf}},
  \bibinfo {author} {\bibfnamefont {J.~S.}\ \bibnamefont {Tsai}}, \ and\
  \bibinfo {author} {\bibfnamefont {O.~V.}\ \bibnamefont {Astafiev}},\ }\href
  {\doibase 10.1038/ncomms12588} {\bibfield  {journal} {\bibinfo  {journal}
  {Nature Communications}\ }\textbf {\bibinfo {volume} {7}},\ \bibinfo {pages}
  {12588} (\bibinfo {year} {2016})}\BibitemShut {NoStop}%
\bibitem [{\citenamefont {Chen}\ \emph {et~al.}(2011)\citenamefont {Chen},
  \citenamefont {Hover}, \citenamefont {Sendelbach}, \citenamefont {Maurer},
  \citenamefont {Merkel}, \citenamefont {Pritchett}, \citenamefont {Wilhelm},\
  and\ \citenamefont {McDermott}}]{chen2011microwave}%
  \BibitemOpen
  \bibfield  {author} {\bibinfo {author} {\bibfnamefont {Y.-F.}\ \bibnamefont
  {Chen}}, \bibinfo {author} {\bibfnamefont {D.}~\bibnamefont {Hover}},
  \bibinfo {author} {\bibfnamefont {S.}~\bibnamefont {Sendelbach}}, \bibinfo
  {author} {\bibfnamefont {L.}~\bibnamefont {Maurer}}, \bibinfo {author}
  {\bibfnamefont {S.~T.}\ \bibnamefont {Merkel}}, \bibinfo {author}
  {\bibfnamefont {E.~J.}\ \bibnamefont {Pritchett}}, \bibinfo {author}
  {\bibfnamefont {F.~K.}\ \bibnamefont {Wilhelm}}, \ and\ \bibinfo {author}
  {\bibfnamefont {R.}~\bibnamefont {McDermott}},\ }\href {\doibase
  10.1103/PhysRevLett.107.217401} {\bibfield  {journal} {\bibinfo  {journal}
  {Physical Review Letters}\ }\textbf {\bibinfo {volume} {107}},\ \bibinfo
  {pages} {217401} (\bibinfo {year} {2011})}\BibitemShut {NoStop}%
\bibitem [{\citenamefont {Zueco}\ \emph {et~al.}(2009)\citenamefont {Zueco},
  \citenamefont {Reuther}, \citenamefont {Kohler},\ and\ \citenamefont
  {H\"anggi}}]{zueco2009qubit}%
  \BibitemOpen
  \bibfield  {author} {\bibinfo {author} {\bibfnamefont {D.}~\bibnamefont
  {Zueco}}, \bibinfo {author} {\bibfnamefont {G.~M.}\ \bibnamefont {Reuther}},
  \bibinfo {author} {\bibfnamefont {S.}~\bibnamefont {Kohler}}, \ and\ \bibinfo
  {author} {\bibfnamefont {P.}~\bibnamefont {H\"anggi}},\ }\href {\doibase
  10.1103/PhysRevA.80.033846} {\bibfield  {journal} {\bibinfo  {journal} {Phys.
  Rev. A}\ }\textbf {\bibinfo {volume} {80}},\ \bibinfo {pages} {033846}
  (\bibinfo {year} {2009})}\BibitemShut {NoStop}%
\bibitem [{\citenamefont {Beaudoin}\ \emph {et~al.}(2011)\citenamefont
  {Beaudoin}, \citenamefont {Gambetta},\ and\ \citenamefont
  {Blais}}]{beadoin2011dissipation}%
  \BibitemOpen
  \bibfield  {author} {\bibinfo {author} {\bibfnamefont {F.}~\bibnamefont
  {Beaudoin}}, \bibinfo {author} {\bibfnamefont {J.~M.}\ \bibnamefont
  {Gambetta}}, \ and\ \bibinfo {author} {\bibfnamefont {A.}~\bibnamefont
  {Blais}},\ }\href {\doibase 10.1103/PhysRevA.84.043832} {\bibfield  {journal}
  {\bibinfo  {journal} {Phys. Rev. A}\ }\textbf {\bibinfo {volume} {84}},\
  \bibinfo {pages} {043832} (\bibinfo {year} {2011})}\BibitemShut {NoStop}%
\bibitem [{\citenamefont {Forn-D\'{\i}az}\ \emph {et~al.}(2010)\citenamefont
  {Forn-D\'{\i}az}, \citenamefont {Lisenfeld}, \citenamefont {Marcos},
  \citenamefont {Garc\'{\i}a-Ripoll}, \citenamefont {Solano}, \citenamefont
  {Harmans},\ and\ \citenamefont {Mooij}}]{forndiaz2010observation}%
  \BibitemOpen
  \bibfield  {author} {\bibinfo {author} {\bibfnamefont {P.}~\bibnamefont
  {Forn-D\'{\i}az}}, \bibinfo {author} {\bibfnamefont {J.}~\bibnamefont
  {Lisenfeld}}, \bibinfo {author} {\bibfnamefont {D.}~\bibnamefont {Marcos}},
  \bibinfo {author} {\bibfnamefont {J.~J.}\ \bibnamefont {Garc\'{\i}a-Ripoll}},
  \bibinfo {author} {\bibfnamefont {E.}~\bibnamefont {Solano}}, \bibinfo
  {author} {\bibfnamefont {C.~J. P.~M.}\ \bibnamefont {Harmans}}, \ and\
  \bibinfo {author} {\bibfnamefont {J.~E.}\ \bibnamefont {Mooij}},\ }\href
  {\doibase 10.1103/PhysRevLett.105.237001} {\bibfield  {journal} {\bibinfo
  {journal} {Phys. Rev. Lett.}\ }\textbf {\bibinfo {volume} {105}},\ \bibinfo
  {pages} {237001} (\bibinfo {year} {2010})}\BibitemShut {NoStop}%
\bibitem [{\citenamefont {Shen}\ and\ \citenamefont
  {Fan}(2009)}]{shen2009theory}%
  \BibitemOpen
  \bibfield  {author} {\bibinfo {author} {\bibfnamefont {J.-T.}\ \bibnamefont
  {Shen}}\ and\ \bibinfo {author} {\bibfnamefont {S.}~\bibnamefont {Fan}},\
  }\href {\doibase 10.1103/PhysRevA.79.023837} {\bibfield  {journal} {\bibinfo
  {journal} {Phys. Rev. A}\ }\textbf {\bibinfo {volume} {79}},\ \bibinfo
  {pages} {023837} (\bibinfo {year} {2009})}\BibitemShut {NoStop}%
\bibitem [{\citenamefont {Rephaeli}\ and\ \citenamefont
  {Fan}(2012)}]{rephaeli2012few}%
  \BibitemOpen
  \bibfield  {author} {\bibinfo {author} {\bibfnamefont {E.}~\bibnamefont
  {Rephaeli}}\ and\ \bibinfo {author} {\bibfnamefont {S.}~\bibnamefont {Fan}},\
  }\href {\doibase 10.1109/JSTQE.2012.2196261} {\bibfield  {journal} {\bibinfo
  {journal} {IEEE J. Sel. Top. Quantum Electron.}\ }\textbf {\bibinfo {volume}
  {18}},\ \bibinfo {pages} {1754} (\bibinfo {year} {2012})}\BibitemShut
  {NoStop}%
\bibitem [{\citenamefont {Oehri}\ \emph {et~al.}(2015)\citenamefont {Oehri},
  \citenamefont {Pletyukhov}, \citenamefont {Gritsev}, \citenamefont
  {Blatter},\ and\ \citenamefont {Schmidt}}]{oehri2015tunable}%
  \BibitemOpen
  \bibfield  {author} {\bibinfo {author} {\bibfnamefont {D.}~\bibnamefont
  {Oehri}}, \bibinfo {author} {\bibfnamefont {M.}~\bibnamefont {Pletyukhov}},
  \bibinfo {author} {\bibfnamefont {V.}~\bibnamefont {Gritsev}}, \bibinfo
  {author} {\bibfnamefont {G.}~\bibnamefont {Blatter}}, \ and\ \bibinfo
  {author} {\bibfnamefont {S.}~\bibnamefont {Schmidt}},\ }\href {\doibase
  10.1103/PhysRevA.91.033816} {\bibfield  {journal} {\bibinfo  {journal} {Phys.
  Rev. A}\ }\textbf {\bibinfo {volume} {91}},\ \bibinfo {pages} {033816}
  (\bibinfo {year} {2015})}\BibitemShut {NoStop}%
\bibitem [{\citenamefont {Hu}\ \emph {et~al.}(2018)\citenamefont {Hu},
  \citenamefont {Zou},\ and\ \citenamefont {Zhang}}]{hu2018}%
  \BibitemOpen
  \bibfield  {author} {\bibinfo {author} {\bibfnamefont {Q.}~\bibnamefont
  {Hu}}, \bibinfo {author} {\bibfnamefont {B.}~\bibnamefont {Zou}}, \ and\
  \bibinfo {author} {\bibfnamefont {Y.}~\bibnamefont {Zhang}},\ }\href
  {\doibase 10.1103/PhysRevA.97.033847} {\bibfield  {journal} {\bibinfo
  {journal} {Phys. Rev. A}\ }\textbf {\bibinfo {volume} {97}},\ \bibinfo
  {pages} {033847} (\bibinfo {year} {2018})}\BibitemShut {NoStop}%
\bibitem [{\citenamefont {Stolyarov}(2019)}]{stolyarov2019few}%
  \BibitemOpen
  \bibfield  {author} {\bibinfo {author} {\bibfnamefont {E.~V.}\ \bibnamefont
  {Stolyarov}},\ }\href {\doibase 10.1103/PhysRevA.99.023857} {\bibfield
  {journal} {\bibinfo  {journal} {Phys. Rev. A}\ }\textbf {\bibinfo {volume}
  {99}},\ \bibinfo {pages} {023857} (\bibinfo {year} {2019})}\BibitemShut
  {NoStop}%
\bibitem [{\citenamefont {Chumak}\ and\ \citenamefont
  {Stolyarov}(2013)}]{chumak2013phase}%
  \BibitemOpen
  \bibfield  {author} {\bibinfo {author} {\bibfnamefont {O.~O.}\ \bibnamefont
  {Chumak}}\ and\ \bibinfo {author} {\bibfnamefont {E.~V.}\ \bibnamefont
  {Stolyarov}},\ }\href {\doibase 10.1103/PhysRevA.88.013855} {\bibfield
  {journal} {\bibinfo  {journal} {Physical Review A}\ }\textbf {\bibinfo
  {volume} {88}},\ \bibinfo {pages} {013855} (\bibinfo {year}
  {2013})}\BibitemShut {NoStop}%
\bibitem [{\citenamefont {Berman}\ and\ \citenamefont
  {Chumak}(2011)}]{berman2011influence}%
  \BibitemOpen
  \bibfield  {author} {\bibinfo {author} {\bibfnamefont {G.~P.}\ \bibnamefont
  {Berman}}\ and\ \bibinfo {author} {\bibfnamefont {A.~A.}\ \bibnamefont
  {Chumak}},\ }\href {\doibase 10.1103/PhysRevA.83.042322} {\bibfield
  {journal} {\bibinfo  {journal} {Physical Review A}\ }\textbf {\bibinfo
  {volume} {83}},\ \bibinfo {pages} {042322} (\bibinfo {year}
  {2011})}\BibitemShut {NoStop}%
\bibitem [{\citenamefont {Braginsky}\ and\ \citenamefont
  {Khalili}(1996)}]{braginsky1996quantum}%
  \BibitemOpen
  \bibfield  {author} {\bibinfo {author} {\bibfnamefont {V.~B.}\ \bibnamefont
  {Braginsky}}\ and\ \bibinfo {author} {\bibfnamefont {F.~Y.}\ \bibnamefont
  {Khalili}},\ }\href {\doibase 10.1103/RevModPhys.68.1} {\bibfield  {journal}
  {\bibinfo  {journal} {Reviews of Modern Physics}\ }\textbf {\bibinfo {volume}
  {68}},\ \bibinfo {pages} {1} (\bibinfo {year} {1996})}\BibitemShut {NoStop}%
\bibitem [{\citenamefont {Bamba}\ and\ \citenamefont
  {Ogawa}(2014)}]{bambaogawa2014recipe}%
  \BibitemOpen
  \bibfield  {author} {\bibinfo {author} {\bibfnamefont {M.}~\bibnamefont
  {Bamba}}\ and\ \bibinfo {author} {\bibfnamefont {T.}~\bibnamefont {Ogawa}},\
  }\href {\doibase 10.1103/PhysRevA.89.023817} {\bibfield  {journal} {\bibinfo
  {journal} {Phys. Rev. A}\ }\textbf {\bibinfo {volume} {89}},\ \bibinfo
  {pages} {023817} (\bibinfo {year} {2014})}\BibitemShut {NoStop}%
\bibitem [{\citenamefont {Boissonneault}\ \emph {et~al.}(2009)\citenamefont
  {Boissonneault}, \citenamefont {Gambetta},\ and\ \citenamefont
  {Blais}}]{boissoneault2009dispersive}%
  \BibitemOpen
  \bibfield  {author} {\bibinfo {author} {\bibfnamefont {M.}~\bibnamefont
  {Boissonneault}}, \bibinfo {author} {\bibfnamefont {J.~M.}\ \bibnamefont
  {Gambetta}}, \ and\ \bibinfo {author} {\bibfnamefont {A.}~\bibnamefont
  {Blais}},\ }\href {\doibase 10.1103/PhysRevA.79.013819} {\bibfield  {journal}
  {\bibinfo  {journal} {Phys. Rev. A}\ }\textbf {\bibinfo {volume} {79}},\
  \bibinfo {pages} {013819} (\bibinfo {year} {2009})}\BibitemShut {NoStop}%
\bibitem [{\citenamefont {Chumak}\ and\ \citenamefont
  {Sushkova}(2012)}]{chumak2012operator}%
  \BibitemOpen
  \bibfield  {author} {\bibinfo {author} {\bibfnamefont {O.}~\bibnamefont
  {Chumak}}\ and\ \bibinfo {author} {\bibfnamefont {N.}~\bibnamefont
  {Sushkova}},\ }\href
  {http://ujp.bitp.kiev.ua/files/journals/57/1/570104p.pdf} {\bibfield
  {journal} {\bibinfo  {journal} {Ukrainian Journal of Physics}\ }\textbf
  {\bibinfo {volume} {57}},\ \bibinfo {pages} {30} (\bibinfo {year}
  {2012})}\BibitemShut {NoStop}%
\bibitem [{\citenamefont {Forn-D{\'i}az}\ \emph {et~al.}(2016)\citenamefont
  {Forn-D{\'i}az}, \citenamefont {Garc{\'i}a-Ripoll}, \citenamefont
  {Peropadre}, \citenamefont {Orgiazzi}, \citenamefont {Yurtalan},
  \citenamefont {Belyansky}, \citenamefont {Wilson},\ and\ \citenamefont
  {Lupascu}}]{forndiaz2016ultrastrong}%
  \BibitemOpen
  \bibfield  {author} {\bibinfo {author} {\bibfnamefont {P.}~\bibnamefont
  {Forn-D{\'i}az}}, \bibinfo {author} {\bibfnamefont {J.~J.}\ \bibnamefont
  {Garc{\'i}a-Ripoll}}, \bibinfo {author} {\bibfnamefont {B.}~\bibnamefont
  {Peropadre}}, \bibinfo {author} {\bibfnamefont {J.-L.}\ \bibnamefont
  {Orgiazzi}}, \bibinfo {author} {\bibfnamefont {M.~A.}\ \bibnamefont
  {Yurtalan}}, \bibinfo {author} {\bibfnamefont {R.}~\bibnamefont {Belyansky}},
  \bibinfo {author} {\bibfnamefont {C.~M.}\ \bibnamefont {Wilson}}, \ and\
  \bibinfo {author} {\bibfnamefont {A.}~\bibnamefont {Lupascu}},\ }\href
  {\doibase 10.1038/nphys3905} {\bibfield  {journal} {\bibinfo  {journal}
  {Nature Physics}\ }\textbf {\bibinfo {volume} {13}},\ \bibinfo {pages} {39}
  (\bibinfo {year} {2016})}\BibitemShut {NoStop}%
\bibitem [{\citenamefont {Gambetta}\ \emph {et~al.}(2007)\citenamefont
  {Gambetta}, \citenamefont {Braff}, \citenamefont {Wallraff}, \citenamefont
  {Girvin},\ and\ \citenamefont {Schoelkopf}}]{gambetta2007protocols}%
  \BibitemOpen
  \bibfield  {author} {\bibinfo {author} {\bibfnamefont {J.}~\bibnamefont
  {Gambetta}}, \bibinfo {author} {\bibfnamefont {W.~A.}\ \bibnamefont {Braff}},
  \bibinfo {author} {\bibfnamefont {A.}~\bibnamefont {Wallraff}}, \bibinfo
  {author} {\bibfnamefont {S.~M.}\ \bibnamefont {Girvin}}, \ and\ \bibinfo
  {author} {\bibfnamefont {R.~J.}\ \bibnamefont {Schoelkopf}},\ }\href@noop {}
  {\bibfield  {journal} {\bibinfo  {journal} {Physical Review A}\ }\textbf
  {\bibinfo {volume} {76}},\ \bibinfo {pages} {012325} (\bibinfo {year}
  {2007})}\BibitemShut {NoStop}%
\bibitem [{\citenamefont {Johnson}\ \emph {et~al.}(2011)\citenamefont
  {Johnson}, \citenamefont {Hoskinson}, \citenamefont {Macklin}, \citenamefont
  {Slichter}, \citenamefont {Siddiqi},\ and\ \citenamefont
  {Clarke}}]{johnson2011dispersive}%
  \BibitemOpen
  \bibfield  {author} {\bibinfo {author} {\bibfnamefont {J.~E.}\ \bibnamefont
  {Johnson}}, \bibinfo {author} {\bibfnamefont {E.~M.}\ \bibnamefont
  {Hoskinson}}, \bibinfo {author} {\bibfnamefont {C.}~\bibnamefont {Macklin}},
  \bibinfo {author} {\bibfnamefont {D.~H.}\ \bibnamefont {Slichter}}, \bibinfo
  {author} {\bibfnamefont {I.}~\bibnamefont {Siddiqi}}, \ and\ \bibinfo
  {author} {\bibfnamefont {J.}~\bibnamefont {Clarke}},\ }\href {\doibase
  10.1103/PhysRevB.84.220503} {\bibfield  {journal} {\bibinfo  {journal} {Phys.
  Rev. B}\ }\textbf {\bibinfo {volume} {84}},\ \bibinfo {pages} {220503(R)}
  (\bibinfo {year} {2011})}\BibitemShut {NoStop}%
\bibitem [{Note1()}]{Note1}%
  \BibitemOpen
  \bibinfo {note} {In our previous work~\cite {sokolov2016optimal} the quantity
  we call here the measurement contrast was erroneously claimed to equal the
  probability of correct measurement result.}\BibitemShut {Stop}%
\bibitem [{\citenamefont {Fan}\ \emph {et~al.}(2014)\citenamefont {Fan},
  \citenamefont {Johansson}, \citenamefont {Combes}, \citenamefont {Milburn},\
  and\ \citenamefont {Stace}}]{fan2014nonabsorbing}%
  \BibitemOpen
  \bibfield  {author} {\bibinfo {author} {\bibfnamefont {B.}~\bibnamefont
  {Fan}}, \bibinfo {author} {\bibfnamefont {G.}~\bibnamefont {Johansson}},
  \bibinfo {author} {\bibfnamefont {J.}~\bibnamefont {Combes}}, \bibinfo
  {author} {\bibfnamefont {G.~J.}\ \bibnamefont {Milburn}}, \ and\ \bibinfo
  {author} {\bibfnamefont {T.~M.}\ \bibnamefont {Stace}},\ }\href {\doibase
  10.1103/PhysRevB.90.035132} {\bibfield  {journal} {\bibinfo  {journal}
  {Physical Review B}\ }\textbf {\bibinfo {volume} {90}},\ \bibinfo {pages}
  {035132} (\bibinfo {year} {2014})}\BibitemShut {NoStop}%
\bibitem [{\citenamefont {Schöndorf}\ \emph {et~al.}(2018)\citenamefont
  {Schöndorf}, \citenamefont {Govia}, \citenamefont {Vavilov}, \citenamefont
  {McDermott},\ and\ \citenamefont {Wilhelm}}]{schoendorf2018optimizing}%
  \BibitemOpen
  \bibfield  {author} {\bibinfo {author} {\bibfnamefont {M.}~\bibnamefont
  {Schöndorf}}, \bibinfo {author} {\bibfnamefont {L.~C.~G.}\ \bibnamefont
  {Govia}}, \bibinfo {author} {\bibfnamefont {M.~G.}\ \bibnamefont {Vavilov}},
  \bibinfo {author} {\bibfnamefont {R.}~\bibnamefont {McDermott}}, \ and\
  \bibinfo {author} {\bibfnamefont {F.~K.}\ \bibnamefont {Wilhelm}},\ }\href
  {\doibase 10.1088/2058-9565/aaa7f7} {\bibfield  {journal} {\bibinfo
  {journal} {Quantum Science and Technology}\ }\textbf {\bibinfo {volume}
  {3}},\ \bibinfo {pages} {024009} (\bibinfo {year} {2018})}\BibitemShut
  {NoStop}%
\bibitem [{\citenamefont {Berman}\ \emph {et~al.}(2013)\citenamefont {Berman},
  \citenamefont {Chumak},\ and\ \citenamefont
  {Tsifrinovich}}]{berman2012dynamics}%
  \BibitemOpen
  \bibfield  {author} {\bibinfo {author} {\bibfnamefont {G.~P.}\ \bibnamefont
  {Berman}}, \bibinfo {author} {\bibfnamefont {A.~A.}\ \bibnamefont {Chumak}},
  \ and\ \bibinfo {author} {\bibfnamefont {V.~I.}\ \bibnamefont
  {Tsifrinovich}},\ }\href {\doibase 10.1007/s10909-012-0726-0} {\bibfield
  {journal} {\bibinfo  {journal} {Journal of Low Temperature Physics}\ }\textbf
  {\bibinfo {volume} {170}},\ \bibinfo {pages} {172} (\bibinfo {year}
  {2013})}\BibitemShut {NoStop}%
\bibitem [{\citenamefont {Sete}\ \emph {et~al.}(2014)\citenamefont {Sete},
  \citenamefont {Gambetta},\ and\ \citenamefont {Korotkov}}]{sete2014purcell}%
  \BibitemOpen
  \bibfield  {author} {\bibinfo {author} {\bibfnamefont {E.~A.}\ \bibnamefont
  {Sete}}, \bibinfo {author} {\bibfnamefont {J.~M.}\ \bibnamefont {Gambetta}},
  \ and\ \bibinfo {author} {\bibfnamefont {A.~N.}\ \bibnamefont {Korotkov}},\
  }\href {\doibase 10.1103/PhysRevB.89.104516} {\bibfield  {journal} {\bibinfo
  {journal} {Physical Review B}\ }\textbf {\bibinfo {volume} {89}},\ \bibinfo
  {pages} {104516} (\bibinfo {year} {2014})}\BibitemShut {NoStop}%
\bibitem [{\citenamefont {Nesterov}\ \emph {et~al.}(2019)\citenamefont
  {Nesterov}, \citenamefont {Pechenezhskiy},\ and\ \citenamefont
  {Vavilov}}]{nesterov2019counting}%
  \BibitemOpen
  \bibfield  {author} {\bibinfo {author} {\bibfnamefont {K.~N.}\ \bibnamefont
  {Nesterov}}, \bibinfo {author} {\bibfnamefont {I.~V.}\ \bibnamefont
  {Pechenezhskiy}}, \ and\ \bibinfo {author} {\bibfnamefont {M.~G.}\
  \bibnamefont {Vavilov}},\ }\href {https://arxiv.org/abs/1911.05933}
  {\bibfield  {journal} {\bibinfo  {journal} {arXiv preprint arXiv:1911.05933}\
  } (\bibinfo {year} {2019})}\BibitemShut {NoStop}%
\bibitem [{\citenamefont {Rigetti}\ \emph {et~al.}(2012)\citenamefont
  {Rigetti}, \citenamefont {Gambetta}, \citenamefont {Poletto}, \citenamefont
  {Plourde}, \citenamefont {Chow}, \citenamefont {C\'orcoles}, \citenamefont
  {Smolin}, \citenamefont {Merkel}, \citenamefont {Rozen}, \citenamefont
  {Keefe}, \citenamefont {Rothwell}, \citenamefont {Ketchen},\ and\
  \citenamefont {Steffen}}]{rigetti2012superconducting}%
  \BibitemOpen
  \bibfield  {author} {\bibinfo {author} {\bibfnamefont {C.}~\bibnamefont
  {Rigetti}}, \bibinfo {author} {\bibfnamefont {J.~M.}\ \bibnamefont
  {Gambetta}}, \bibinfo {author} {\bibfnamefont {S.}~\bibnamefont {Poletto}},
  \bibinfo {author} {\bibfnamefont {B.~L.~T.}\ \bibnamefont {Plourde}},
  \bibinfo {author} {\bibfnamefont {J.~M.}\ \bibnamefont {Chow}}, \bibinfo
  {author} {\bibfnamefont {A.~D.}\ \bibnamefont {C\'orcoles}}, \bibinfo
  {author} {\bibfnamefont {J.~A.}\ \bibnamefont {Smolin}}, \bibinfo {author}
  {\bibfnamefont {S.~T.}\ \bibnamefont {Merkel}}, \bibinfo {author}
  {\bibfnamefont {J.~R.}\ \bibnamefont {Rozen}}, \bibinfo {author}
  {\bibfnamefont {G.~A.}\ \bibnamefont {Keefe}}, \bibinfo {author}
  {\bibfnamefont {M.~B.}\ \bibnamefont {Rothwell}}, \bibinfo {author}
  {\bibfnamefont {M.~B.}\ \bibnamefont {Ketchen}}, \ and\ \bibinfo {author}
  {\bibfnamefont {M.}~\bibnamefont {Steffen}},\ }\href {\doibase
  10.1103/PhysRevB.86.100506} {\bibfield  {journal} {\bibinfo  {journal} {Phys.
  Rev. B}\ }\textbf {\bibinfo {volume} {86}},\ \bibinfo {pages} {100506(R)}
  (\bibinfo {year} {2012})}\BibitemShut {NoStop}%
\bibitem [{\citenamefont {Paik}\ \emph {et~al.}(2016)\citenamefont {Paik},
  \citenamefont {Mezzacapo}, \citenamefont {Sandberg}, \citenamefont {McClure},
  \citenamefont {Abdo}, \citenamefont {C\'orcoles}, \citenamefont {Dial},
  \citenamefont {Bogorin}, \citenamefont {Plourde}, \citenamefont {Steffen},
  \citenamefont {Cross}, \citenamefont {Gambetta},\ and\ \citenamefont
  {Chow}}]{paik2016experimental}%
  \BibitemOpen
  \bibfield  {author} {\bibinfo {author} {\bibfnamefont {H.}~\bibnamefont
  {Paik}}, \bibinfo {author} {\bibfnamefont {A.}~\bibnamefont {Mezzacapo}},
  \bibinfo {author} {\bibfnamefont {M.}~\bibnamefont {Sandberg}}, \bibinfo
  {author} {\bibfnamefont {D.~T.}\ \bibnamefont {McClure}}, \bibinfo {author}
  {\bibfnamefont {B.}~\bibnamefont {Abdo}}, \bibinfo {author} {\bibfnamefont
  {A.~D.}\ \bibnamefont {C\'orcoles}}, \bibinfo {author} {\bibfnamefont
  {O.}~\bibnamefont {Dial}}, \bibinfo {author} {\bibfnamefont {D.~F.}\
  \bibnamefont {Bogorin}}, \bibinfo {author} {\bibfnamefont {B.~L.~T.}\
  \bibnamefont {Plourde}}, \bibinfo {author} {\bibfnamefont {M.}~\bibnamefont
  {Steffen}}, \bibinfo {author} {\bibfnamefont {A.~W.}\ \bibnamefont {Cross}},
  \bibinfo {author} {\bibfnamefont {J.~M.}\ \bibnamefont {Gambetta}}, \ and\
  \bibinfo {author} {\bibfnamefont {J.~M.}\ \bibnamefont {Chow}},\ }\href
  {\doibase 10.1103/PhysRevLett.117.250502} {\bibfield  {journal} {\bibinfo
  {journal} {Phys. Rev. Lett.}\ }\textbf {\bibinfo {volume} {117}},\ \bibinfo
  {pages} {250502} (\bibinfo {year} {2016})}\BibitemShut {NoStop}%
\bibitem [{\citenamefont {Yan}\ \emph {et~al.}(2016)\citenamefont {Yan},
  \citenamefont {Gustavsson}, \citenamefont {Kamal}, \citenamefont {Birenbaum},
  \citenamefont {Sears}, \citenamefont {Hover}, \citenamefont {Gudmundsen},
  \citenamefont {Rosenberg}, \citenamefont {Samach}, \citenamefont {Weber},
  \citenamefont {Yoder}, \citenamefont {Orlando}, \citenamefont {Clarke},
  \citenamefont {Kerman},\ and\ \citenamefont {Oliver}}]{yan2016fluxrevisited}%
  \BibitemOpen
  \bibfield  {author} {\bibinfo {author} {\bibfnamefont {F.}~\bibnamefont
  {Yan}}, \bibinfo {author} {\bibfnamefont {S.}~\bibnamefont {Gustavsson}},
  \bibinfo {author} {\bibfnamefont {A.}~\bibnamefont {Kamal}}, \bibinfo
  {author} {\bibfnamefont {J.}~\bibnamefont {Birenbaum}}, \bibinfo {author}
  {\bibfnamefont {A.~P.}\ \bibnamefont {Sears}}, \bibinfo {author}
  {\bibfnamefont {D.}~\bibnamefont {Hover}}, \bibinfo {author} {\bibfnamefont
  {T.~J.}\ \bibnamefont {Gudmundsen}}, \bibinfo {author} {\bibfnamefont
  {D.}~\bibnamefont {Rosenberg}}, \bibinfo {author} {\bibfnamefont
  {G.}~\bibnamefont {Samach}}, \bibinfo {author} {\bibfnamefont
  {S.}~\bibnamefont {Weber}}, \bibinfo {author} {\bibfnamefont {J.~L.}\
  \bibnamefont {Yoder}}, \bibinfo {author} {\bibfnamefont {T.~P.}\ \bibnamefont
  {Orlando}}, \bibinfo {author} {\bibfnamefont {J.}~\bibnamefont {Clarke}},
  \bibinfo {author} {\bibfnamefont {A.~J.}\ \bibnamefont {Kerman}}, \ and\
  \bibinfo {author} {\bibfnamefont {W.~D.}\ \bibnamefont {Oliver}},\ }\href
  {\doibase 10.1038/ncomms12964} {\bibfield  {journal} {\bibinfo  {journal}
  {Nature Communications}\ }\textbf {\bibinfo {volume} {7}},\ \bibinfo {pages}
  {12964} (\bibinfo {year} {2016})}\BibitemShut {NoStop}%
\bibitem [{\citenamefont {Jeffrey}\ \emph {et~al.}(2014)\citenamefont
  {Jeffrey}, \citenamefont {Sank}, \citenamefont {Mutus}, \citenamefont
  {White}, \citenamefont {Kelly}, \citenamefont {Barends}, \citenamefont
  {Chen}, \citenamefont {Chen}, \citenamefont {Chiaro}, \citenamefont
  {Dunsworth}, \citenamefont {Megrant}, \citenamefont {O'Malley}, \citenamefont
  {Neill}, \citenamefont {Roushan}, \citenamefont {Vainsencher}, \citenamefont
  {Wenner}, \citenamefont {Cleland},\ and\ \citenamefont
  {Martinis}}]{jeffrey2014fast}%
  \BibitemOpen
  \bibfield  {author} {\bibinfo {author} {\bibfnamefont {E.}~\bibnamefont
  {Jeffrey}}, \bibinfo {author} {\bibfnamefont {D.}~\bibnamefont {Sank}},
  \bibinfo {author} {\bibfnamefont {J.~Y.}\ \bibnamefont {Mutus}}, \bibinfo
  {author} {\bibfnamefont {T.~C.}\ \bibnamefont {White}}, \bibinfo {author}
  {\bibfnamefont {J.}~\bibnamefont {Kelly}}, \bibinfo {author} {\bibfnamefont
  {R.}~\bibnamefont {Barends}}, \bibinfo {author} {\bibfnamefont
  {Y.}~\bibnamefont {Chen}}, \bibinfo {author} {\bibfnamefont {Z.}~\bibnamefont
  {Chen}}, \bibinfo {author} {\bibfnamefont {B.}~\bibnamefont {Chiaro}},
  \bibinfo {author} {\bibfnamefont {A.}~\bibnamefont {Dunsworth}}, \bibinfo
  {author} {\bibfnamefont {A.}~\bibnamefont {Megrant}}, \bibinfo {author}
  {\bibfnamefont {P.~J.~J.}\ \bibnamefont {O'Malley}}, \bibinfo {author}
  {\bibfnamefont {C.}~\bibnamefont {Neill}}, \bibinfo {author} {\bibfnamefont
  {P.}~\bibnamefont {Roushan}}, \bibinfo {author} {\bibfnamefont
  {A.}~\bibnamefont {Vainsencher}}, \bibinfo {author} {\bibfnamefont
  {J.}~\bibnamefont {Wenner}}, \bibinfo {author} {\bibfnamefont {A.~N.}\
  \bibnamefont {Cleland}}, \ and\ \bibinfo {author} {\bibfnamefont {J.~M.}\
  \bibnamefont {Martinis}},\ }\href {\doibase 10.1103/PhysRevLett.112.190504}
  {\bibfield  {journal} {\bibinfo  {journal} {Phys. Rev. Lett.}\ }\textbf
  {\bibinfo {volume} {112}},\ \bibinfo {pages} {190504} (\bibinfo {year}
  {2014})}\BibitemShut {NoStop}%
\bibitem [{\citenamefont {Chapman}\ \emph {et~al.}(2017)\citenamefont
  {Chapman}, \citenamefont {Rosenthal}, \citenamefont {Kerckhoff},
  \citenamefont {Moores}, \citenamefont {Vale}, \citenamefont {Mates},
  \citenamefont {Hilton}, \citenamefont {Lalumi\`ere}, \citenamefont {Blais},\
  and\ \citenamefont {Lehnert}}]{chapman2017widely}%
  \BibitemOpen
  \bibfield  {author} {\bibinfo {author} {\bibfnamefont {B.~J.}\ \bibnamefont
  {Chapman}}, \bibinfo {author} {\bibfnamefont {E.~I.}\ \bibnamefont
  {Rosenthal}}, \bibinfo {author} {\bibfnamefont {J.}~\bibnamefont
  {Kerckhoff}}, \bibinfo {author} {\bibfnamefont {B.~A.}\ \bibnamefont
  {Moores}}, \bibinfo {author} {\bibfnamefont {L.~R.}\ \bibnamefont {Vale}},
  \bibinfo {author} {\bibfnamefont {J.~A.~B.}\ \bibnamefont {Mates}}, \bibinfo
  {author} {\bibfnamefont {G.~C.}\ \bibnamefont {Hilton}}, \bibinfo {author}
  {\bibfnamefont {K.}~\bibnamefont {Lalumi\`ere}}, \bibinfo {author}
  {\bibfnamefont {A.}~\bibnamefont {Blais}}, \ and\ \bibinfo {author}
  {\bibfnamefont {K.~W.}\ \bibnamefont {Lehnert}},\ }\href {\doibase
  10.1103/PhysRevX.7.041043} {\bibfield  {journal} {\bibinfo  {journal} {Phys.
  Rev. X}\ }\textbf {\bibinfo {volume} {7}},\ \bibinfo {pages} {041043}
  (\bibinfo {year} {2017})}\BibitemShut {NoStop}%
\bibitem [{\citenamefont {Mahoney}\ \emph {et~al.}(2017)\citenamefont
  {Mahoney}, \citenamefont {Colless}, \citenamefont {Pauka}, \citenamefont
  {Hornibrook}, \citenamefont {Watson}, \citenamefont {Gardner}, \citenamefont
  {Manfra}, \citenamefont {Doherty},\ and\ \citenamefont
  {Reilly}}]{mahoney2017onchip}%
  \BibitemOpen
  \bibfield  {author} {\bibinfo {author} {\bibfnamefont {A.~C.}\ \bibnamefont
  {Mahoney}}, \bibinfo {author} {\bibfnamefont {J.~I.}\ \bibnamefont
  {Colless}}, \bibinfo {author} {\bibfnamefont {S.~J.}\ \bibnamefont {Pauka}},
  \bibinfo {author} {\bibfnamefont {J.~M.}\ \bibnamefont {Hornibrook}},
  \bibinfo {author} {\bibfnamefont {J.~D.}\ \bibnamefont {Watson}}, \bibinfo
  {author} {\bibfnamefont {G.~C.}\ \bibnamefont {Gardner}}, \bibinfo {author}
  {\bibfnamefont {M.~J.}\ \bibnamefont {Manfra}}, \bibinfo {author}
  {\bibfnamefont {A.~C.}\ \bibnamefont {Doherty}}, \ and\ \bibinfo {author}
  {\bibfnamefont {D.~J.}\ \bibnamefont {Reilly}},\ }\href {\doibase
  10.1103/PhysRevX.7.011007} {\bibfield  {journal} {\bibinfo  {journal} {Phys.
  Rev. X}\ }\textbf {\bibinfo {volume} {7}},\ \bibinfo {pages} {011007}
  (\bibinfo {year} {2017})}\BibitemShut {NoStop}%
\bibitem [{\citenamefont {Bernier}\ \emph {et~al.}(2017)\citenamefont
  {Bernier}, \citenamefont {T{\'o}th}, \citenamefont {Koottandavida},
  \citenamefont {Ioannou}, \citenamefont {Malz}, \citenamefont {Nunnenkamp},
  \citenamefont {Feofanov},\ and\ \citenamefont
  {Kippenberg}}]{bernier2017nonreciprocal}%
  \BibitemOpen
  \bibfield  {author} {\bibinfo {author} {\bibfnamefont {N.~R.}\ \bibnamefont
  {Bernier}}, \bibinfo {author} {\bibfnamefont {L.~D.}\ \bibnamefont
  {T{\'o}th}}, \bibinfo {author} {\bibfnamefont {A.}~\bibnamefont
  {Koottandavida}}, \bibinfo {author} {\bibfnamefont {M.~A.}\ \bibnamefont
  {Ioannou}}, \bibinfo {author} {\bibfnamefont {D.}~\bibnamefont {Malz}},
  \bibinfo {author} {\bibfnamefont {A.}~\bibnamefont {Nunnenkamp}}, \bibinfo
  {author} {\bibfnamefont {A.~K.}\ \bibnamefont {Feofanov}}, \ and\ \bibinfo
  {author} {\bibfnamefont {T.~J.}\ \bibnamefont {Kippenberg}},\ }\href
  {\doibase 10.1038/s41467-017-00447-1} {\bibfield  {journal} {\bibinfo
  {journal} {Nature Communications}\ }\textbf {\bibinfo {volume} {8}},\
  \bibinfo {pages} {604} (\bibinfo {year} {2017})}\BibitemShut {NoStop}%
\bibitem [{\citenamefont {Forn-D\'{\i}az}\ \emph {et~al.}(2017)\citenamefont
  {Forn-D\'{\i}az}, \citenamefont {Warren}, \citenamefont {Chang},
  \citenamefont {Vadiraj},\ and\ \citenamefont {Wilson}}]{forndiaz2017on}%
  \BibitemOpen
  \bibfield  {author} {\bibinfo {author} {\bibfnamefont {P.}~\bibnamefont
  {Forn-D\'{\i}az}}, \bibinfo {author} {\bibfnamefont {C.~W.}\ \bibnamefont
  {Warren}}, \bibinfo {author} {\bibfnamefont {C.~W.~S.}\ \bibnamefont
  {Chang}}, \bibinfo {author} {\bibfnamefont {A.~M.}\ \bibnamefont {Vadiraj}},
  \ and\ \bibinfo {author} {\bibfnamefont {C.~M.}\ \bibnamefont {Wilson}},\
  }\href {\doibase 10.1103/PhysRevApplied.8.054015} {\bibfield  {journal}
  {\bibinfo  {journal} {Phys. Rev. Applied}\ }\textbf {\bibinfo {volume} {8}},\
  \bibinfo {pages} {054015} (\bibinfo {year} {2017})}\BibitemShut {NoStop}%
\bibitem [{\citenamefont {Oelsner}\ and\ \citenamefont
  {Il’ichev}(2018)}]{oelsner2018switching}%
  \BibitemOpen
  \bibfield  {author} {\bibinfo {author} {\bibfnamefont {G.}~\bibnamefont
  {Oelsner}}\ and\ \bibinfo {author} {\bibfnamefont {E.}~\bibnamefont
  {Il’ichev}},\ }\href {\doibase 10.1007/s10909-018-1959-3} {\bibfield
  {journal} {\bibinfo  {journal} {Journal of Low Temperature Physics}\ }\textbf
  {\bibinfo {volume} {192}},\ \bibinfo {pages} {169} (\bibinfo {year}
  {2018})}\BibitemShut {NoStop}%
\bibitem [{\citenamefont {Inomata}\ \emph {et~al.}(2016)\citenamefont
  {Inomata}, \citenamefont {Lin}, \citenamefont {Koshino}, \citenamefont
  {Oliver}, \citenamefont {Tsai}, \citenamefont {Yamamoto},\ and\ \citenamefont
  {Nakamura}}]{inomata2016single}%
  \BibitemOpen
  \bibfield  {author} {\bibinfo {author} {\bibfnamefont {K.}~\bibnamefont
  {Inomata}}, \bibinfo {author} {\bibfnamefont {Z.}~\bibnamefont {Lin}},
  \bibinfo {author} {\bibfnamefont {K.}~\bibnamefont {Koshino}}, \bibinfo
  {author} {\bibfnamefont {W.~D.}\ \bibnamefont {Oliver}}, \bibinfo {author}
  {\bibfnamefont {J.-S.}\ \bibnamefont {Tsai}}, \bibinfo {author}
  {\bibfnamefont {T.}~\bibnamefont {Yamamoto}}, \ and\ \bibinfo {author}
  {\bibfnamefont {Y.}~\bibnamefont {Nakamura}},\ }\href
  {https://doi.org/10.1038/ncomms12303} {\bibfield  {journal} {\bibinfo
  {journal} {Nature Communications}\ }\textbf {\bibinfo {volume} {7}},\
  \bibinfo {pages} {12303} (\bibinfo {year} {2016})}\BibitemShut {NoStop}%
\bibitem [{\citenamefont {Besse}\ \emph {et~al.}(2018)\citenamefont {Besse},
  \citenamefont {Gasparinetti}, \citenamefont {Collodo}, \citenamefont
  {Walter}, \citenamefont {Kurpiers}, \citenamefont {Pechal}, \citenamefont
  {Eichler},\ and\ \citenamefont {Wallraff}}]{besse2018singleshot}%
  \BibitemOpen
  \bibfield  {author} {\bibinfo {author} {\bibfnamefont {J.-C.}\ \bibnamefont
  {Besse}}, \bibinfo {author} {\bibfnamefont {S.}~\bibnamefont {Gasparinetti}},
  \bibinfo {author} {\bibfnamefont {M.~C.}\ \bibnamefont {Collodo}}, \bibinfo
  {author} {\bibfnamefont {T.}~\bibnamefont {Walter}}, \bibinfo {author}
  {\bibfnamefont {P.}~\bibnamefont {Kurpiers}}, \bibinfo {author}
  {\bibfnamefont {M.}~\bibnamefont {Pechal}}, \bibinfo {author} {\bibfnamefont
  {C.}~\bibnamefont {Eichler}}, \ and\ \bibinfo {author} {\bibfnamefont
  {A.}~\bibnamefont {Wallraff}},\ }\href {\doibase 10.1103/PhysRevX.8.021003}
  {\bibfield  {journal} {\bibinfo  {journal} {Phys. Rev. X}\ }\textbf {\bibinfo
  {volume} {8}},\ \bibinfo {pages} {021003} (\bibinfo {year}
  {2018})}\BibitemShut {NoStop}%
\bibitem [{\citenamefont {Kaplunenko}\ and\ \citenamefont
  {Ustinov}(2004)}]{kaplunenko2004experimental}%
  \BibitemOpen
  \bibfield  {author} {\bibinfo {author} {\bibfnamefont {V.~K.}\ \bibnamefont
  {Kaplunenko}}\ and\ \bibinfo {author} {\bibfnamefont {A.~V.}\ \bibnamefont
  {Ustinov}},\ }\href {\doibase 10.1140/epjb/e2004-00091-3} {\bibfield
  {journal} {\bibinfo  {journal} {Eur. Phys. J. B}\ }\textbf {\bibinfo {volume}
  {38}},\ \bibinfo {pages} {3} (\bibinfo {year} {2004})}\BibitemShut {NoStop}%
\bibitem [{\citenamefont {Liebermann}\ and\ \citenamefont
  {Wilhelm}(2016)}]{liebermann2016optimal}%
  \BibitemOpen
  \bibfield  {author} {\bibinfo {author} {\bibfnamefont {P.~J.}\ \bibnamefont
  {Liebermann}}\ and\ \bibinfo {author} {\bibfnamefont {F.~K.}\ \bibnamefont
  {Wilhelm}},\ }\href {\doibase 10.1103/PhysRevApplied.6.024022} {\bibfield
  {journal} {\bibinfo  {journal} {Phys. Rev. Applied}\ }\textbf {\bibinfo
  {volume} {6}},\ \bibinfo {pages} {024022} (\bibinfo {year}
  {2016})}\BibitemShut {NoStop}%
\bibitem [{\citenamefont {Klenov}\ \emph {et~al.}(2017)\citenamefont {Klenov},
  \citenamefont {Kuznetsov}, \citenamefont {Soloviev}, \citenamefont
  {Bakurskiy}, \citenamefont {Denisenko},\ and\ \citenamefont
  {Satanin}}]{klenov2017flux}%
  \BibitemOpen
  \bibfield  {author} {\bibinfo {author} {\bibfnamefont {N.~V.}\ \bibnamefont
  {Klenov}}, \bibinfo {author} {\bibfnamefont {A.~V.}\ \bibnamefont
  {Kuznetsov}}, \bibinfo {author} {\bibfnamefont {I.~I.}\ \bibnamefont
  {Soloviev}}, \bibinfo {author} {\bibfnamefont {S.~V.}\ \bibnamefont
  {Bakurskiy}}, \bibinfo {author} {\bibfnamefont {M.~V.}\ \bibnamefont
  {Denisenko}}, \ and\ \bibinfo {author} {\bibfnamefont {A.~M.}\ \bibnamefont
  {Satanin}},\ }\href {\doibase 10.1063/1.4995627} {\bibfield  {journal}
  {\bibinfo  {journal} {Low Temperature Physics}\ }\textbf {\bibinfo {volume}
  {43}},\ \bibinfo {pages} {789} (\bibinfo {year} {2017})}\BibitemShut
  {NoStop}%
\bibitem [{\citenamefont {McDermott}\ and\ \citenamefont
  {Vavilov}(2014)}]{mcdermott2014accurate}%
  \BibitemOpen
  \bibfield  {author} {\bibinfo {author} {\bibfnamefont {R.}~\bibnamefont
  {McDermott}}\ and\ \bibinfo {author} {\bibfnamefont {M.~G.}\ \bibnamefont
  {Vavilov}},\ }\href {\doibase 10.1103/PhysRevApplied.2.014007} {\bibfield
  {journal} {\bibinfo  {journal} {Phys. Rev. Applied}\ }\textbf {\bibinfo
  {volume} {2}},\ \bibinfo {pages} {014007} (\bibinfo {year}
  {2014})}\BibitemShut {NoStop}%
\end{thebibliography}%
